\newif\ifpreprint

\preprintfalse 

\ifpreprint
\documentclass[aip,jcp,amsmath,amssymb,preprint,longbibliography]{revtex4-1}
\else
\documentclass[aip,jcp,amsmath,amssymb,reprint,longbibliography]{revtex4-1}
\fi

\usepackage[utf8]{inputenc}

\usepackage{appendix}
\usepackage{amsmath}
\usepackage{graphicx}
\usepackage{threeparttable}
\usepackage{ifthen}
\usepackage[table]{xcolor}
\usepackage{xspace}
\usepackage{xr}
\usepackage[version=4]{mhchem}
\usepackage{lineno}
\usepackage{braket}
\usepackage{placeins}
\usepackage{gensymb}
\usepackage{multirow}
\usepackage[colorlinks = false]{hyperref}
\usepackage[capitalise]{cleveref}
\usepackage{hhline}
\usepackage{braket}

\definecolor{goodorange}{RGB}{225,125,0}
\definecolor{goodgreen}{RGB}{0,125,0}
\definecolor{goodred}{RGB}{220,50,25}
\definecolor{goodblue}{RGB}{25,25,150}

\newcommand{\note}[2]{
\ifthenelse{\equal{#1}{F}}{
\colorbox{goodorange}{\textcolor{white}{\footnotesize \fontfamily{phv}\selectfont #1}}
    \textcolor{goodorange}{{\footnotesize \fontfamily{phv}\selectfont #2}}\xspace
}{}
\ifthenelse{\equal{#1}{S}}{
\colorbox{goodblue}{\textcolor{white}{\footnotesize \fontfamily{phv}\selectfont #1}}
    \textcolor{goodblue}{{\footnotesize \fontfamily{phv}\selectfont #2}}\xspace
}{}
\ifthenelse{\equal{#1}{Z}}{
\colorbox{goodgreen}{\textcolor{white}{\footnotesize \fontfamily{phv}\selectfont #1}}
    \textcolor{goodgreen}{{\footnotesize \fontfamily{phv}\selectfont #2}}\xspace
}{}
}

\newcommand{\cas}[0]{\hat{\mathcal{R}}^{\textrm{int}}_\mathrm{CAS}}
\newcommand{\single}[0]{\hat{\mathcal{R}}^{\textrm{int}}_\text{pMBS}}
\newcommand{\double}[0]{\hat{\mathcal{R}}^{\textrm{int}}_\text{pMBD}}

\modulolinenumbers[5]

\begin{document}

\title{Equation-of-motion internally contracted multireference unitary coupled-cluster theory}

\author{Shuhang Li}
\email{shuhang.li@emory.edu}
\author{Zijun Zhao}
\author{Francesco A. Evangelista}
\email{francesco.evangelista@emory.edu}
\affiliation{Department of Chemistry and Cherry Emerson Center for Scientific Computation, Emory University, Atlanta, Georgia 30322, United States}

\begin{abstract}
The accurate computation of excited states remains a challenge in electronic structure theory, especially for systems with a ground state that requires a multireference treatment. 
In this work, we introduce a novel equation-of-motion (EOM) extension of the internally contracted multireference unitary coupled-cluster framework (ic-MRUCC), termed EOM-ic-MRUCC. 
EOM-ic-MRUCC follows the transform-then-diagonalize approach, in analogy to its non-unitary counterpart [Datta and Nooijen, \textit{J. Chem. Phys.} \textbf{137}, 204107 (2012)]. 
By employing a projective approach to optimize the ground state, the method retains additive separability and proper scaling with system size.
We show that excitation energies are size intensive if the EOM operator satisfies the ``killer'' and the projective conditions.
Furthermore, we propose to represent changes in reference state upon electron excitation via projected many-body operators that span active orbitals and show that the EOM equations formulated in this way are invariant with respect to active orbital rotations.
We test the EOM-ic-MRUCC method truncated to single and double excitations by computing the potential energy curves for several excited states of a \ce{BeH2} model system, the HF molecule, and water undergoing symmetric dissociation.
 Across these systems, our method delivers accurate excitation energies and potential energy curves within 5 m$E_\mathrm{h}$ (ca. 0.14 eV) from full configuration interaction. We find that truncating the Baker--Campbell--Hausdorff series to four-fold commutators contributes negligible errors (on the order of $10^{-5}$ $E_\mathrm{h}$ or less), offering a practical route to highly accurate excited-state calculations with reduced computational overhead.
\end{abstract}
\maketitle
\pagebreak

\section{Introduction}
The accurate description of the electronic structure of molecules in excited states remains a long-standing challenge in quantum chemistry, primarily due to the intricate interplay of static and dynamical electron correlation effects.
Most well-established excited-state methods fall under the single-reference (SR) category and assume that the ground state is dominated by a single configuration, often a closed-shell Hartree--Fock determinant.
Owing to their computational efficiency and simplicity, configuration interaction with single excitations (CIS)~\cite{foresman1992toward} and time-dependent density functional theory (TD-DFT)~\cite{casida1995time, furche2002adiabatic} are among the most popular SR approaches.
Excited-state methods based on coupled-cluster (CC) theory~\cite{Cizek:1966cy,jankowski.1991.10.1007/BF01117411,piecuch.1999.10.1063/1.478517,Crawford:2000by,Bartlett:2007kv} include CC linear response (CC-LR)~\cite{monkhorst1977calculation, mukherjee1979response, dalgaard1983some, koch1990coupled} and equation-of-motion CC theory (EOM-CC),\cite{rowe.1968.10.1103/RevModPhys.40.153, geertsen1989equation, stanton1993equation, mukhopadhyay1991aspects, nooijen1995description,watts1996iterative} which are equivalent as far as excitation energies are concerned.~\cite{sokolov2018multi}
A closely related formalism, the symmetry-adapted cluster configuration interaction (SAC-CI), was also proposed by Nakatsuji.~\cite{nakatsuji1979cluster}
Furthermore, CC methods that directly target individual states have been proposed.\cite{meissner.1993.10.1016/0009-26149387127-O,jankowski.1994.10.1002/qua.560500504,jankowski.1995.10.1002/qua.560530507,kowalski.1998.10.1103/PhysRevLett.81.1195,jankowski.1999.10.1063/1.478900,jankowski.1999.10.1063/1.479575,piecuch.2000.10.1142/9789812792501_0001,mayhall.2010.10.1021/ct100321k,lee.2019.10.1063/1.5128795,kossoski.2021.10.1021/acs.jctc.1c00348,damour.2024.10.1021/acs.jctc.4c00034}
Excited-state CC methods are the preferred choice for high-accuracy computations due to their conceptual simplicity and ability to accurately capture excitations while retaining size-extensivity and size-intensivity properties.~\cite{hanrath2009concepts}
While CC methods and their EOM extension can be applied to ground states that are not well described by a single determinant (\textit{e.g.}, at transition state geometries, bond-recoupling regions, and open-shell transition metal complexes), in practice, the performance of conventional truncation schemes deteriorates significantly.
Several solutions to this problem have been proposed, including: the use of multi-ionization and attachment EOM manifolds,\cite{musial.2011.10.1063/1.3567115,musial.2020.10.1002/9781119417774.ch4} adding a smaller subset of higher excitations in the cluster operator,\cite{kowalski.2000.10.1063/1.1318757,kowalski.2001.10.1063/1.1378323,piecuch.2010.10.1080/00268976.2010.522608} method-of-moments,\cite{piecuch.2004.10.1007/s00214-004-0567-2,loch.2006.10.1080/00268970600659586,gururangan.2025.10.1016/j.cplett.2024.141840} and spin-flip formulations of EOM-CC.\cite{krylov.2006.10.1021/ar0402006}

Multireference CC (MRCC) methods\cite{piecuch.1992.10.1007/BF01113244,paldus.1993.10.1103/PhysRevA.47.2738,piecuch.1993.10.1063/1.466179,piecuch.1994.10.1063/1.467304,piecuch.1994.10.1063/1.467143,hubavc1994size, mavsik1998single, mahapatra.1998.10.1016/S0065-32760860507-9, mahapatra.1998.10.1080/002689798168448, mahapatra1999size, mahapatra2010potential, mahapatra2011evaluation,lyakh.2012.10.1021/cr2001417} that start from a correlated reference wavefunction provide the basis for extending EOM-CC to the multireference domain.
Jagau \textit{et al.}~\cite{jagau2012linear, jagau2012linear2} reported a linear response formalism based on Mukherjee's MRCC theory (Mk-MRCC-LR).
Because this approach is based on the Jeziorski--Monkhorst ansatz,\cite{jeziorski.1981.10.1103/PhysRevA.24.1668} the number of amplitudes scales linearly with the number of reference determinants, leading to exponential scaling with the number of active orbitals.
At the same time, Mk-MRCC-LR excited-state energies also depend on the choice of active orbitals since the ground-state energy is not invariant with respect to active-active orbital rotations.
Internally contracted MRCC (ic-MRCC) and related formalisms~\cite{mukherjee1975correlation, banerjee1981coupled, banerjee1982applications, evangelista2011orbital, hanauer2011pilot, hanauer2012communication, datta2011state, yanai2006canonical, yanai2007canonical, feldmann2024renormalized, mazziotti2006anti, mazziotti2007anti, greenman2008electronic} address these issues by expressing the ground state as a single exponential operator acting on a linear combination of reference determinants.
Excited-state methods based on variants of internally contracted formalisms include the MR-EOMCC approach developed by Nooijen and co-workers~\cite{datta2012multireference} and linear response ic-MRCC (ic-MRCC-LR) introduced by Samanta \textit{et al.}~\cite{samanta2014excited} 
The reliance of excited-state MRCC methods on non-Hermitian effective Hamiltonians is problematic as it can potentially result in unphysical complex eigenvalues.~\cite{piecuch1990coupled,kowalski.2000.10.1002/1097-461X200080:4/5<757::AID-QUA25>3.0.CO;2-A,datta2012multireference}
This problem is also observed in SR EOM-CC methods.\cite{kohn.2007.10.1063/1.2755681,kjonstad.2017.10.1063/1.4998724, thomas2021complex}

The goal of this paper is to examine excited-state methods based on \textit{unitary} MRCC (MRUCC).
Unitary formulations of CC theory (UCC)\cite{kutzelnigg.1977.10.1007/978-1-4757-0887-5_5,kucharski1991hilbert,kutzelnigg.1983.10.1063/1.446313,kutzelnigg.1984.10.1063/1.446736,bartlett.1989.10.1016/S0009-26148987372-5,szalay.1995.10.1063/1.469641,majee.2024.10.1063/5.0207091} offer notable advantages over traditional CC, such as faster convergence towards full configuration interaction (FCI) and preservation of Hermiticity.\cite{cooper.2010.10.1063/1.3520564,evangelistaAlternativeSinglereferenceCoupled2011, harsha.2018.10.1063/1.5011033,hodecker.2020.10.1063/1.5142354,tolle.2023.10.1063/5.0139716}
Excited-state formulations of UCC\cite{hodecker.2020.10.1063/5.0019055,liu.2018.10.1063/1.5030344,liu.2021.10.1063/5.0062090} lead to a consistent calculation of properties via either the sum-over-states polarization propagator and response theory,\cite{taubeNewPerspectivesUnitary2006} and approximate variants of UCC have been shown to be connected to the GW method.\cite{tolle.2023.10.1063/5.0139716}
UCC was originally introduced in the early days of quantum chemistry but remained mostly an academic curiosity since it leads to a non-terminating Baker--Campbell--Hausdorff (BCH) expansion of the similarity-transformed Hamiltonian.\cite{kutzelnigg1991error}
In this direction, a related work is the multireference algebraic diagrammatic construction (MR-ADC) theory developed by Sokolov and co-workers for the simulation of various spectroscopic processes.~\cite{sokolov2018multi, mazin2021multireference, chatterjee2019second, chatterjee2020extended, mazin2023core, de2024efficient}
MR-ADC is a Hermitian multireference propagator method, in which the ground state is formally equivalent to ic-MRUCC. Practical MR-ADC truncation schemes treat both ground and excited states at the perturbative level, with the highest level treatment achieved to date being the extended second order.\cite{mazin2021multireference}

Recently, renewed interest in UCC has stemmed from the observation by Peruzzo \textit{et al.}~\cite{peruzzo2014variational} that UCC realized as a product of unitary operators can be implemented on a quantum computer in polynomial time using a series of gate-based operations.
Various quantum computing approaches have been proposed to approximate both ground and excited state molecular energies based on UCC,~\cite{aspuru-guzik_simulated_2005, mcclean_hybrid_2017, motta_determining_2020, stair_simulating_2021, smart_quantum_2021,benavides-riveros_quantum_2024, otten_localized_2022, xie_power_2022, wen_full_2024,tkachenko_quantum_2024,grimsley.2024.10.48550/arXiv.2409.11210,ollitrault.2020.10.1103/PhysRevResearch.2.043140,asthanaQuantumSelfconsistentEquationofmotion2023,kjellgren.2024.10.1063/5.0225409,ziems.2024.10.48550/arXiv.2408.09308} many of which are quantum variants of classical EOM or LR approaches.
Several of these methods may be considered forms of EOM multireference UCC.
For example, qEOM~\cite{ollitrault.2020.10.1103/PhysRevResearch.2.043140} solves a set of EOM equations where the ground state is approximated with a UCC state with singles and doubles (UCCSD).
The q-sc-EOM~\cite{asthanaQuantumSelfconsistentEquationofmotion2023,kimTwoAlgorithmsExcitedState2023} and quantum linear response (qLR) formalisms\cite{kumar.2023.10.1021/acs.jctc.3c00731} adopted the self-consistent EOM operator formulation,\cite{prasad1985some, datta1993consistent} and proposed a projective operator variant.
In both cases, a factorized UCC ground state was built adaptively starting from a Hartree--Fock reference.\cite{grimsley.2019.10.1038/s41467-019-10988-2}
While the abovementioned papers considered all orbitals in the system, a series of recent works investigated qLR with excitation expansion truncated to an active space in combination with orbital response in the full orbital space.\cite{kjellgren.2024.10.1063/5.0225409,ziems.2024.10.48550/arXiv.2408.09308,ziems2024options,jensen.2024.10.1021/acs.jctc.4c00069,reinholdt.2024.10.1021/acs.jctc.4c00211,vonbuchwald.2024.10.1021/acs.jctc.4c00574}
 This may be viewed as an MCSCF linear response formalism in which the reference is an orbital-optimized UCCSD state spanning only the active orbitals.\cite{benavides-riveros.2024.10.1088/1367-2630/ad2d1d,wang.2023.10.1103/PhysRevA.108.022814}

In this work, we present an EOM extension of the ic-MRUCC approach based on a correlated reference generated within the conventional active-space partitioning of the orbitals.
We adopt a projective approach to determine the ground-state ic-MRUCC amplitudes, \cite{evangelistaAlternativeSinglereferenceCoupled2011} which, unlike variational formulations,\cite{chen2012orbitally} is both size consistent and size extensive.
Previous quantum computing implementations of EOM-UCC use a product form of the UCC state to capture electron correlation from either all orbitals or only the active orbitals.
In contrast, our approach explicitly separates the contributions from the active orbitals, treating the remaining correlation effects in the ground and excited states with operators that promote the electrons to and from at least one orbital outside the active space.
These operators are responsible for both orbital relaxation and two-body correlation effects.
This development is also designed to inform EOM extensions of approximate UCC-based theories.
We are specifically interested in the multireference driven similarity renormalization group (MR-DSRG) formalism,~\cite{evangelista2014driven, li2015multireference, li2017driven, li2016towards, li2019multireference} an approximate, renormalized variant of ic-MRUCC robust to numerical instabilities.

The remainder of the paper is organized as follows.
\cref{sec:theory} details the projective formulation of ic-MRUCC theory and the strategies used to eliminate operator linear dependencies. This section also presents the EOM extension of ic-MRUCC, discusses the conditions required to achieve size-intensive excitation energies, and describes our parameterization of the EOM excitation operator.
\cref{sec:implementation} provides details of our EOM-ic-MRUCC pilot implementation.
In \cref{sec:results}, we report benchmark results comparing EOM-ic-MRUCC calculations to FCI results for three model systems and examine how truncating the BCH expansion affects accuracy.
Finally, in \cref{sec:conclusion}, we summarize our findings and highlight future research directions, including potential extensions to approximate UCC-based theories within the DSRG framework.

\section{Theory}
\label{sec:theory}
\subsection{Notation}
To establish our formalism, we begin by introducing the notation and conventions employed throughout this work.
We consider a multideterminantal reference state, $\ket{\Phi_0}$, obtained from a complete-active-space self-consistent-field (CASSCF) computation. 
This state is a linear combination of Slater determinants $\ket{\phi_{\mu}}$ weighted by normalized coefficients $c_{\mu}$:
\begin{align}
	\ket{\Phi_0} = \sum_{\mu = 1}^{d}\ket{\phi_\mu} c_{\mu}. \label{eq:casscf}
\end{align}
The set of determinants $M = \{ \ket{\phi_\mu}, \mu =1, \ldots, d \}$ defines the model space, and it is constructed from a set of spin orbitals partitioned into core (\textbf{C}), active (\textbf{A}), and virtual (\textbf{V}) subsets of sizes $N_{\textbf{C}}$, $N_{\textbf{A}}$, and $N_{\textbf{V}}$, respectively.
The notation CAS($m$e, $n$o) denotes a complete active space with $m$ electrons distributed among $n$ spatial active orbitals.
We use the indices \textit{m}, \textit{n} to designate core spin orbitals, \textit{u}, \textit{v}, \textit{w}, \textit{x}, \textit{y}, \textit{z}, \textit{t} to represent active spin orbitals, and \textit{e}, \textit{f}, \textit{g}, \textit{h} for virtual spin orbitals. 
We also introduce two composite orbital subsets: the hole spin orbitals ($\textbf{H} = \textbf{C} \cup \textbf{A}$) and the particle spin orbitals ($\textbf{P} = \textbf{A} \cup \textbf{V}$) of dimensions $N_{\textbf{H}} = N_{\textbf{C}} + N_{\textbf{A}}$ and $N_{\textbf{P}} = N_{\textbf{A}} + N_{\textbf{V}}$, respectively. 
The hole spin orbitals are denoted by the labels \textit{i}, \textit{j}, \textit{k}, \textit{l}, while the particle spin orbitals are labeled with \textit{a}, \textit{b}, \textit{c}, \textit{d}. 
General spin orbitals that belong to either the hole or particle set are designated as \textit{p}, \textit{q}, \textit{r}, \textit{s}.

\subsection{Internally contracted multireference unitary coupled-cluster theory}
\label{sec:icmrucc}

The ic-MRUCC ansatz parameterizes the exact many-body ground state with a unitary exponential operator, $e^{\hat{A}}$, acting on the general reference state $\ket{\Phi_0}$:
\begin{equation}
\ket{\Psi_0} = e^{\hat{A}}\ket{\Phi_0} = e^{\hat{T} - \hat{T}^{\dagger}}\ket{\Phi_0}. \label{eq:ic-mrucc}
\end{equation}
The operator $\hat{A} = \hat{T} - \hat{T}^\dagger$ is the anti-Hermitian combination of the cluster operator $\hat{T}$. The operator $\hat{T}$ may include hole-particle excitation operators with up to $n$-fold substitutions, where $n$ is the number of electrons in the system.

To express the ic-MRUCC equations in a compact form, we introduce general amplitudes ($t_q \equiv t_{ab\cdots}^{ij\cdots}$) and excitation operators ($\hat{\tau}_q  \equiv \{ \hat{a}^{ab\cdots}_{ij\cdots} \}$) identified by a collective index $q = ({ij\cdots},{ab\cdots})$ that ranges from one to the total number of excitation operators ($n_{\mathrm{ex}}$):
\begin{equation}
	\hat{T} = \sum_{q = 1}^{n_{\mathrm{ex}}}t_q\hat{\tau}_q.
\end{equation}
Brackets ($\{ \cdot \}$) are used to indicate normal ordering with respect to the correlated state $\Phi_0$, using the generalized normal-ordering (GNO) formalism.\cite{mukherjeeNormalOrderingWicklike1997, kutzelniggNormalOrderExtended1997, kongAlgebraicProofGeneralized2010}
The cluster amplitudes $t_{ab\cdots}^{ij\cdots}$ in the ic-MRUCC ansatz are tensors antisymmetric with respect to the individual permutation of upper and lower indices. 
Internal amplitudes corresponding to substitutions involving only active orbitals are not included, \textit{i.e.}, $t_{uv\cdots}^{xy\cdots}=0$,~\cite{mahapatra.1998.10.1016/S0065-32760860507-9} as their role is to relax the coefficients of the reference determinants, which is made redundant by the reference relaxation procedure discussed below.\cite{evangelista2011orbital}

Inserting the ic-MRUCC ansatz [see \cref{eq:ic-mrucc}] into the Schr{\"o}dinger equation and left-multiplying by $e^{-\hat{A}}$ leads to
\begin{equation}
	e^{-\hat{A}}\hat{H}e^{\hat{A}}\ket{\Phi_0} = \bar{H}\ket{\Phi_0} = E_0 \ket{\Phi_0}, \label{eq:trans_sch_equation}
\end{equation}
where we have introduced the transformed Hamiltonian $\bar{H} = e^{-\hat{A}} \hat{H} e^{\hat{A}}$.
Cluster amplitude equations are obtained by left-projecting \cref{eq:trans_sch_equation} onto a linearly independent set of internally contracted excited functions.
An obvious choice is the set of internally contracted excited configurations $\{\hat{\tau}_q\ket{\Phi_0}\}$; however, these states are not guaranteed to be linearly independent since the corresponding metric matrix
\begin{equation}
(\mathbf{S})_{pq} = \braket{\Phi_0|\hat{\tau}^{\dagger}_p \hat{\tau}_q|\Phi_0},
\label{eq:mrcc overlap}
\end{equation}
is generally singular.
This linear dependency introduces numerical instabilities, prompting the need to eliminate redundant configurations in most internally contracted MR methods.~\cite{lowdin1970nonorthogonality, werner1988efficient, andersson1990second, evangelista2011orbital, hanauer2011pilot, hanauer2012communication}

Following earlier implementations of ic-MRCC,~\cite{evangelista2011orbital, hanauer2011pilot, hanauer2012communication} we address the issue of linear dependencies by expressing $\hat{T}$ in terms of linearly independent operators $\hat{\kappa}_Q$ and corresponding amplitudes $k_Q$ as:
\begin{equation}
	\hat{T} = \sum_{Q = 1}^{n_{\mathrm{in}}}k_Q\hat{\kappa}_Q,
\end{equation}
where $n_{\mathrm{in}}$ is the number of linearly independent operators.
Writing the linearly independent excitation operators as a row vector, $\hat{\boldsymbol{\kappa}} = (\hat{\kappa}_1,\ldots,\hat{\kappa}_{n_\mathrm{in}})$, we can express them in terms of the original excitation operators [$\hat{\boldsymbol{\tau}} = (\hat{\tau}_1, \ldots, \hat{\tau}_{n_\mathrm{ex}} )$] via the rectangular matrix $\mathbf{X}$ as:
\begin{equation}
\label{eq:ktx}
\hat{\boldsymbol{\kappa}} = \hat{\boldsymbol{\tau}}\mathbf{X}.
\end{equation}
Various methods exist for defining $\mathbf{X}$, and prior studies have determined ways to ensure the size extensivity of the corresponding ic-MRCC theory.~\cite{hanauer2012communication}
These strategies can also be extended to formulate a size-extensive ic-MRUCC theory, as discussed in \cref{sec:size-extensive}.
What is important to note is that since we define ic-MRUCC using normal-ordered operators, size extensivity is guaranteed irrespective of the orthogonalization procedure used to obtain $\mathbf{X}$.

Projecting \cref{eq:trans_sch_equation} onto the set of linearly independent internally contracted states leads to the following set of amplitude equations:
\begin{equation}
\braket{\Phi_0|\hat{\kappa}_Q^{\dagger}\bar{H}|\Phi_0} = 0.
\label{eq:orthogonal amplitude equations}
\end{equation}
The energy and expansion coefficients are instead obtained by projecting \cref{eq:trans_sch_equation} onto the model space determinants, resulting in the following eigenvalue equation
\begin{equation}
	\sum_{\nu}\braket{\phi_{\mu}|\bar{H}|\phi_{\nu}} c_{\nu} = E_0 c_{\mu}. \label{eq:effective hamiltonian}
\end{equation}
To relax the reference state in the presence of dynamical correlation effects included via the exponential operator, \cref{eq:orthogonal amplitude equations,{eq:effective hamiltonian}} must be solved iteratively until self-consistency is reached.
This is also a necessary condition for the ic-MRUCC state to be exact.

In concluding this section, we discuss the exactness of the ic-MRUCC ansatz in relation to ic-MRCC and single-reference UCC theory.
In the case of ic-MRCC, the convergence of the wavefunction toward the full configuration interaction limit as the maximum excitation rank increases was proved by constructing the corresponding exact cluster operator $\hat{T}$.~\cite{evangelista2012sequential}
Extending such proof to the case of a unitary operator runs into difficulties already encountered in the case of single-reference UCC, where numerical evidence has been unable to find counterexamples to exactness, but a formal proof of exactness is unknown.~\cite{evangelista2019exact}
Therefore, at present, the exactness of the ic-MRUCC ansatz can only be assumed as a working hypothesis and tested numerically.

\subsection{Equation-of-motion extension of ic-MRUCC}
\label{sec:eom}
Following EOM theory,~\cite{rowe.1968.10.1103/RevModPhys.40.153,geertsen1989equation, stanton1993equation, mukhopadhyay1991aspects, nooijen1995description, watts1996iterative} we define the EOM-ic-MRUCC ansatz as
\begin{equation}
	\ket{\Psi_{\alpha}} = \bar{\mathcal{R}}_{\alpha}\ket{\Psi_0},
\end{equation}
where $\bar{\mathcal{R}}_{\alpha}$ is a formal excitation operator delivering the $\alpha$-th excited state from the ic-MRUCC ground state, \textit{i.e.}, formally $\bar{\mathcal{R}}_{\alpha}\equiv\ket{\Psi_{\alpha}}\bra{\Psi_0}$.
The Schr{\"o}dinger equation for this excited state reads
\begin{equation}
	\hat{H}\bar{\mathcal{R}}_{\alpha}\ket{\Psi_0}=E_{\alpha}\bar{\mathcal{R}}_{\alpha}\ket{\Psi_0},
\end{equation}
whereas the ground-state Schr{\"o}dinger equation, left-multiplied by $\bar{\mathcal{R}}_{\alpha}$, reads
\begin{equation}
	\bar{\mathcal{R}}_{\alpha}\hat{H}\ket{\Psi_0}=E_{0}\bar{\mathcal{R}}_{\alpha}\ket{\Psi_0},
\end{equation}
where $E_{0}$ and $E_{\alpha}$ are the ground- and excited-state energies, respectively.
Taking the difference of these two equations eliminates the ground-state energy and yields:
\begin{equation}
\label{eq:eom-master}
	[\hat{H},\bar{\mathcal{R}}_{\alpha}]\ket{\Psi_0}=\omega_{\alpha}\bar{\mathcal{R}}_{\alpha}\ket{\Psi_0},
\end{equation}
where $\omega_{\alpha}=E_{\alpha}-E_0$ denotes the excitation energy of the $\alpha$-th state. 

Due to its formal advantages, we adopt the ``self-consistent'' excitation operators introduced by Mukherjee and coworkers, \cite{prasad1985some, datta1993consistent} which expresses $\bar{\mathcal{R}}_{\alpha}$ as a similarity-transformed operator:
\begin{equation}
    \bar{\mathcal{R}}_{\alpha}\equiv e^{\hat{A}}\hat{\mathcal{R}}_{\alpha}e^{-\hat{A}}.
\end{equation}
Substituting this expression into \cref{eq:eom-master} and multiplying on the left by $\exp(-\hat{A})$ we arrive at:
\begin{equation}
	[\bar{H},\hat{\mathcal{R}}_{\alpha}]\ket{\Phi_0}=\omega_{\alpha}\hat{\mathcal{R}}_{\alpha}\ket{\Phi_0}.\label{eq:psi_k}
\end{equation}
The advantage of the self-consistent operator approach is that it yields an expression for the excitation energy [\cref{eq:psi_k}] in terms of the similarity-transformed Hamiltonian $\bar{H} = e^{-\hat{A}} \hat{H} e^{\hat{A}}$.

Next, we consider the problem of how we should parameterize the operator $\bar{\mathcal{R}}_{\alpha}$ to make the EOM approach exact.
The excited state generated by $\bar{\mathcal{R}}_{\alpha}$ may be expressed as:
\begin{equation}
\ket{\Psi_{\alpha}} = \bar{\mathcal{R}}_{\alpha} \ket{\Psi_0} = e^{\hat{A}}\hat{\mathcal{R}}_{\alpha}\ket{\Phi_0},
\end{equation}
or equivalently
\begin{equation}
\hat{\mathcal{R}}_{\alpha}\ket{\Phi_0} =  e^{-\hat{A}} \ket{\Psi_{\alpha}}.
\label{eq:Ra_phi_0}
\end{equation}
We may interpret \cref{eq:Ra_phi_0} as a constraint on the form of $\hat{\mathcal{R}}_{\alpha}$: this operator must transform the reference $\ket{\Phi_0}$ into a general vector in Hilbert space.
Additionally, orthogonality between the ground and excited states implies that $\hat{\mathcal{R}}_{\alpha}\ket{\Phi_0}$ must be orthogonal to $\ket{\Phi_0}$ since $\bra{\Phi_0} \hat{\mathcal{R}}_{\alpha}\ket{\Phi_0} = \bra{\Phi_0} e^{-\hat{A}} \ket{\Psi_{\alpha}} = \braket{\Psi_0 | \Psi_\alpha} = 0$.
For convenience, we will refer to the condition:
\begin{equation}
\label{eq:weakkiller}
\bra{\Phi_0} \hat{\mathcal{R}}_{\alpha}\ket{\Phi_0} = 0,
\end{equation}
as the weak form of the killer condition. 
This condition is implied by the stronger killer condition (also known as the vacuum annihilation condition),~\cite{goscinski1980role, weiner1980self, prasad1985some, datta1993consistent, szekeres2001killer} which is often invoked in the formulation of EOM methods:
\begin{equation}
\label{eq:killer}
	\hat{\mathcal{R}}_{\alpha}^\dagger \ket{\Phi_0} = 0.
\end{equation}

When parameterized as a linear operator, $\hat{\mathcal{R}}_{\alpha}$ must be a combination of excitation operators $\{\hat{{\rho}}_p\}$ with corresponding excitation amplitudes $r_{\alpha}^{p}$:
\begin{equation}
\label{eq:r operator}
	\hat{\mathcal{R}}_{\alpha} = \sum_{p = 1}^{n_{\mathrm{eom}}}r_{\alpha}^{p}\hat{\rho}_p.
\end{equation}
Projecting \cref{eq:psi_k} on the left onto the set of internally contracted configurations $\{\hat{\rho}_p\ket{\Phi_0}\}$ and using \cref{eq:r operator} lead to the following generalized eigenvalue problem:
\begin{equation}
	\sum_{q=1}^{n_{\mathrm{eom}}}\braket{\Phi_0|\hat{\rho}_p^{\dagger}[\bar{H},\hat{\rho}_q]|\Phi_0} r_{\alpha}^q
	= \omega_{\alpha} \sum_{q=1}^{n_{\mathrm{eom}}} \braket{\Phi_0|\hat{\rho}_p^{\dagger}\hat{\rho}_q|\Phi_0} r_{\alpha}^q.
\label{eq:eom_eigenvalue_1}
\end{equation}
Orthogonality among the excited states may be expressed as a condition on the inner product of EOM operators:
\begin{equation}
\braket{\Psi_\alpha | \Psi_\beta}
=
\bra{\Phi_0}
\hat{\mathcal{R}}_{\alpha}^\dagger
\hat{\mathcal{R}}_{\beta}\ket{\Phi_0} 
= 
\sum_{p,q=1}^{n_{\mathrm{eom}}} 
r_{\alpha}^p
\braket{\Phi_0|\hat{\rho}_p^{\dagger}\hat{\rho}_q|\Phi_0} r_{\beta}^q
= \delta_{\alpha \beta}.
\end{equation}

If $\hat{\mathcal{R}}_{\alpha}$ satisfies the killer condition [\cref{eq:killer}],
it is possible to rewrite the EOM equation only in terms of commutators as:
\begin{equation}
\label{eq:connected_eom}
	\sum_{q=1}^{n_{\mathrm{eom}}}\braket{\Phi_0|[\hat{\rho}_p^{\dagger},[\bar{H},\hat{\rho}_q]]|\Phi_0} r_{\alpha}^q
	= \omega_{\alpha} \sum_{q=1}^{n_{\mathrm{eom}}} \braket{\Phi_0|[\hat{\rho}_p^{\dagger},\hat{\rho}_q]|\Phi_0} r_{\alpha}^q.
\end{equation}
This second form of the EOM equations is generally preferred to \cref{eq:eom_eigenvalue_1} as it leads to an expansion in terms of fully connected diagrams, ensuring size-intensive excitation energies. Moreover, for a fully relaxed ic-MRUCC state, the matrix entering the left-hand side is Hermitian due to the killer and the projective amplitude conditions.
For unrelaxed ic-MRUCC states, one may formulate a Hermitian eigenvalue problem by symmetrizing the double commutator.

To arrive at a practical EOM-ic-MRUCC scheme, we need to specify a form for the excitation operator $\hat{\mathcal{R}}_{\alpha}$.
This operator may be separated into internal ($\hat{\mathcal{R}}^{\mathrm{int}}_{\alpha}$) and external ($\hat{\mathcal{R}}^{\mathrm{ext}}_{\alpha}$) components:
\begin{equation}
\hat{\mathcal{R}}_{\alpha} = 
\hat{\mathcal{R}}^{\mathrm{int}}_{\alpha}
+ \hat{\mathcal{R}}^{\mathrm{ext}}_{\alpha} ,
\end{equation}
where internal excitations map the model space into itself ($\hat{\mathcal{R}}^{\mathrm{int}}_{\alpha} M \in M$), while external excitations generate excited configurations outside the model space ($\hat{\mathcal{R}}^{\mathrm{ext}}_{\alpha} M \notin M$). 
Operators from these two groups span orthogonal spaces when applied to the state $\ket{\Phi_0}$.
Note that scalar terms should be excluded from $\hat{\mathcal{R}}_{\alpha}$ to ensure that the weaker form of the killer condition, $\braket{\Phi_0 | \hat{\mathcal{R}}_{\alpha} | \Phi_0} = 0$, is satisfied.

We parameterize external excitations with the same set of many-body operators in $\hat{T}$.
For the ic-MRUCCSD method considered in this study, this set comprises all one and two hole-particle substitution operators, excluding those labeled only by active indices.
Omitting the target excited state label ($\alpha$) we can write the external excitations as:
\begin{equation}
\hat{\mathcal{R}}^{\mathrm{ext}} = \hat{\mathcal{R}}^{\mathrm{ext}}_{1} + \hat{\mathcal{R}}^{\mathrm{ext}}_{2},
\end{equation}
where a generic $k$-body term is defined as
\begin{equation}
\hat{\mathcal{R}}^{\mathrm{ext}}_{k}
= \frac{1}{(k\,!)^2}\sum_{ij\cdots}^{\textbf{H}}\sum_{ab\cdots}^{\textbf{P}}r_{ab\cdots}^{ij\cdots}\{ \hat{a}^{ab\cdots}_{ij\cdots} \}.
\end{equation}
When parameterized this way, the adjoint of $\hat{\mathcal{R}}^{\mathrm{ext}}$ contains operators of the form $\{ \hat{a}_{ab\cdots}^{ij\cdots} \}$, which annihilate the reference state.
Therefore, $\hat{\mathcal{R}}^{\mathrm{ext}}$ naturally satisfies the killer condition.

There are several possible parameterizations for internal excitations, leading to different ways to realize EOM-ic-MRUCC.
A common and convenient choice is to represent internal EOM excitations using transfer operators between the ground and excited CASCI solutions:~\cite{lowdin1985some,kutzelnigg1989time,sokolov2018multi, mazin2021multireference} 
\begin{align}
\hat{\mathcal{R}}^{\textrm{int}}_\mathrm{CAS} = \sum_{I>0} r_I \ket{\Phi_I}\bra{\Phi_0}.
\end{align}
This parameterization satisfies the killer condition, but it is impractical for large active spaces due to the combinatorial growth of the number of excited CASCI solutions with the number of electrons and active orbitals.~\cite{sokolov2018multi,mazin2021multireference}
Since the convergence of the excitation energies with the number of transfer operators is generally fast, a practical solution is to work with a small set of CASCI states representing low-lying excited states.
This comes at the cost of compromising orbital invariance in the EOM step, as the truncated CASCI states no longer constitute a complete basis for the model space.

In this work, we propose to use an alternative solution based on projected many-body (pMB) internal excitations.
This approach is analogous to the projected operator set first proposed by Szekeres \textit{et al.},\cite{szekeres2001killer} and then by several recent works on quantum EOM and LR theories.\cite{fan.2021.10.1021/acs.jpclett.1c02153,liu.2022.10.1039/D1CS01184G,kumar.2023.10.1021/acs.jctc.3c00731, ziems2024options,kjellgren.2024.10.1063/5.0225409,ziems.2024.10.48550/arXiv.2408.09308}
To satisfy the killer condition, we start from many-body internal operators ($\{ \hat{a}^{xy\cdots}_{uv\cdots} \}$) defined to contain only active indices.
From the internal operators, we define the corresponding projected form as $\hat{\rho}_q = \{ \hat{a}^{xy\cdots}_{uv\cdots} \} \ket{\Phi_0}\bra{\Phi_0}-\braket{\Phi_0|\{ \hat{a}^{xy\cdots}_{uv\cdots} \}|\Phi_0}$.
Since the operators $\{ \hat{a}^{xy\cdots}_{uv\cdots} \}$ are normal ordered w.r.t.\ $\ket{\Phi_0}$,  the second term vanishes by definition,\cite{mukherjeeNormalOrderingWicklike1997} \textit{i.e.}, $\hat{\rho}_q=\{ \hat{a}^{xy\cdots}_{uv\cdots} \}\ket{\Phi_0}\bra{\Phi_0}$.
Moreover, these new operators satisfy the killer conditions since $\hat{\rho}_q^\dagger \ket{\Phi_0} = \ket{\Phi_0}\braket{\Phi_0 | \{ \hat{a}_{xy\cdots}^{uv\cdots} \} | \Phi_0} = 0$, where in the last term $\braket{\Phi_0 | \{ \hat{a}_{xy\cdots}^{uv\cdots} \} | \Phi_0} = 0$ due to the normal-ordering condition.
In the case of singles and doubles, the projected many-body internals are defined as:
\begin{equation}
\hat{\mathcal{R}}^{\mathrm{int}}_\text{pMBS}
= \sum_{ux}^{\textbf{A}}r_{x}^{u} \{ \hat{a}^{x}_{u} \} \ket{\Phi_0}\bra{\Phi_0},
\end{equation}
and
\begin{equation}
\hat{\mathcal{R}}^{\mathrm{int}}_\text{pMBD}
= \frac{1}{4}\sum_{uvxy}^{\textbf{A}}r_{xy}^{uv}\{ \hat{a}^{xy}_{uv} \} \ket{\Phi_0}\bra{\Phi_0}.
\end{equation}
The main advantage of the pMB formulation is that it reduces the number of variables from combinatorial to polynomial scaling.
Higher order pMB operators can be included in $\hat{\mathcal{R}}^{\mathrm{int}}$ to improve the description of excitations involving active orbitals systematically.
When the space of internal excitations is saturated, the pMB approach becomes equivalent to the transfer operator approach.
In principle, it is also possible to determine eigenstates of $\hat{H}$ within the space spanned by the pMB states ($\{ \hat{a}^{xy\cdots}_{uv\cdots} \} \ket{\Phi_0}$) and then truncate them by keeping a small number of low-energy eigenstates, as done in the transfer operator approach. This approach is not explored in this work.

The matrix representation of the connected EOM-ic-MRUCC equation [\cref{eq:connected_eom}] is:
\begin{equation}
	\bar{\mathbf{H}}' \mathbf{r}_{\alpha} = \omega_{\alpha} \mathbf{S}'\mathbf{r}_{\alpha}, \label{eq:matrix eom}
\end{equation} 
where the matrices $\bar{\mathbf{H}}'$ and $\mathbf{S}'$ are defined as
\begin{equation}
\bar{H}'_{pq} =\braket{\Phi_0|[\hat{\rho}_p^{\dagger},[\bar{H},\hat{\rho}_q]]|\Phi_0},	
\end{equation}
and
\begin{equation}
S'_{pq} =\braket{\Phi_0|[\hat{\rho}_p^{\dagger},\hat{\rho}_q]|\Phi_0}.
\end{equation}
Note that the $\bar{\mathbf{H}}'$ and $\mathbf{S}'$ matrices are constructed using the $\hat{\rho}_p$ operators (either internal or external excitations) and can in principle differ from the matrices $\bar{\mathbf{H}}$ and $\mathbf{S}$ that appear in ic-MRUCC.
Like for the ground-state problem, the set of internally contracted excited configurations $\{\hat{\rho}_p\ket{\Phi_0}\}$ exhibit linear dependence and can be orthogonalized via a linear transformation induced by the matrix $\mathbf{M}$:
\begin{equation}
	\boldsymbol{\hat{\chi}} = \boldsymbol{\hat{\rho}}\mathbf{M}. \label{eq:generalized eigen start}
\end{equation}
When expressed in the $\boldsymbol{\hat{\chi}}$ operator basis, \cref{eq:matrix eom} becomes an ordinary eigenvalue problem:
\begin{equation}
	\bar{\mathbf{H}}'' \mathbf{r}_{\alpha}' = \omega_\alpha \mathbf{r}_{\alpha}',\label{eq:ge3}
\end{equation}
where the modified matrix $\bar{\mathbf{H}}''$ and vector $\mathbf{r'}_{\alpha}$ are defined as	$\mathbf{\bar{H}}'' = \mathbf{M}^{\dagger}\bar{\mathbf{H}}'\mathbf{M}$
and $\mathbf{r}'_{\alpha} = \mathbf{M}^{\dagger}\mathbf{r}_{\alpha}$.

\subsection{Size intensivity of excitation energies and orbital invariance of EOM-ic-MRUCC}
\label{sec:size-intensive}
In this section, we discuss the scaling properties of the EOM-ic-MRUCC excitation energies.
The additive separability of the ground-state energy is instead discussed in \cref{sec:size-extensive}, where we also report numerical tests to confirm that a GNO-based formulation of this approach leads to size-consistent energies.

For the EOM extension of ic-MRUCC, we require that excitation energies remain constant in the presence of non-interacting subsystems (size intensivity) and that they remain unchanged by unitary rotations that separately mix core, active, and virtual orbitals (orbital invariance).
The requirement of size intensivity has implications on the choice of the excitation operator manifold. 
As we prove in \cref{app:intensivity}, the killer condition is necessary for satisfying size intensivity.
As discussed in \cref{sec:eom}, our parameterization of the $\hat{\mathcal{R}}_{\alpha}$ operator automatically satisfies \cref{eq:killer}.
Each operator in the orthogonalized EOM manifold, $\{\hat{\chi}_Q\}$, further needs to satisfy the following condition:
\begin{equation}
\label{eq:chi-dagger}
\braket{\Phi_0|\hat{\chi}_Q^{\dagger}\bar{H}|\Phi_0} = 0.
\end{equation}
To show that $\hat{\mathcal{R}}_{\alpha}$ also satisfies \cref{eq:chi-dagger}, we rewrite the l.h.s.\ by inserting the resolution of the identity in terms of CASCI eigenstates ($\Phi_I$) and linearly independent EOM external excited configurations ($\hat{\chi}_Q \ket{\Phi_0}$):
\begin{equation}
\begin{split}
\braket{\Phi_0|\hat{\chi}_P^{\dagger}\bar{H}|\Phi_0}
& = \sum_I^{\mathrm{CAS}} \braket{\Phi_0|\hat{\chi}_P^{\dagger}|\Phi_I}\braket{\Phi_I|\bar{H}|\Phi_0} \\
&+ \sum_Q^\mathrm{ext} \braket{\Phi_0|\hat{\chi}_P^{\dagger}\hat{\chi}_Q|\Phi_0}\braket{\Phi_0|\hat{\chi}_Q^{\dagger}\bar{H}|\Phi_0}.
\label{eq:chi-dagger-split}
\end{split}
\end{equation}
If the ground state reference is an eigenstate of $\bar{H}$, i.e., it satisfies \cref{eq:effective hamiltonian}, then the first term of \cref{eq:chi-dagger-split} reduces to $\braket{\Phi_0|\hat{\chi}_P^{\dagger}|\Phi_0}\braket{\Phi_0|\bar{H}|\Phi_0}$, and the factor $\braket{\Phi_0|\hat{\chi}_P^{\dagger}|\Phi_0}$ is null since $\hat{\chi}_P^{\dagger}$ is a linear combination of GNO operators.
The second term in \cref{eq:chi-dagger-split} is zero because the factor $\braket{\Phi_0|\hat{\chi}_Q^{\dagger}\bar{H}|\Phi_0}$ is guaranteed to be null by the amplitude equation [\cref{eq:orthogonal amplitude equations}] and the equivalence of the external excitation orthonormalized operator basis of $\hat{T}$ and $\hat{\mathcal{R}}_{\alpha}^{\mathrm{ext}}$.

To emphasize the importance of GNO in the definition of the internal EOM excitations, we conducted a numerical test to assess size intensivity, employing the eigenoperator basis ($\cas$) and the projected many-body singles ($\single$) with and without using normal-ordered operators.
For this test, we use the \ce{LiH} + \ce{H2} composite system and expand the active spaces for \ce{LiH} and the composite system to CAS(2,4) and CAS(4,6), respectively (see \cref{sec:size-extensive}).
The differences in the first triplet excitation energy between the composite system and the \ce{LiH} subsystem were found to be $1.3 \times 10^{-7} E_{\mathrm{h}}$ when employing $\single$ without normal-ordered operators.
This difference was instead less than $10^{-12} E_{\mathrm{h}}$ when employing either the $\cas$ or $\single$ with GNO operators, consistent with our formal analysis.
The GNO transformation is therefore employed for all EOM-ic-MRUCC computations without explicit labeling.

Lastly, we note another advantage of formulating the EOM-ic-MRUCC using projected many-body excitations.
For large active spaces that might require truncation, using pMB operators preserves the orbital invariance property if all elements of a certain excitation type are included, in contrast to a truncated eigenoperator basis that will not satisfy this condition.

\section{Implementation}
\label{sec:implementation}

The EOM-ic-MRUCC method has been implemented in a pilot code interfaced with \textsc{Forte},~\cite{evangelista2024forte} an open-source plugin for the \textsc{Psi4} \emph{ab initio} quantum chemistry package.~\cite{smith2020psi4}
Our implementation leverages \textsc{Forte}'s functionality to represent arbitrary linear combinations of determinants and second-quantized operators using a bit array representation.
A general state $\ket{\Omega}$ is expressed as a linear combination of determinants ($\ket{\phi_\mu}$) as:
\begin{equation}
	\ket{\Omega} = \sum_{\mu}c_{\mu}\ket{\phi_\mu}.
\end{equation}
The action of $e^{\hat{A}}$ on a state vector $\ket{\Omega}$ is computed using the Taylor expansion:
\begin{equation}
	e^{\hat{A}}\ket{\Omega} = \ket{\Omega} + \hat{A}\ket{\Omega} + \frac{1}{2!}\hat{A}^{2}\ket{\Omega} + \cdots,
\end{equation}
and is truncated when the largest absolute element of the vector $\frac{1}{m!}\hat{A}^{m}\ket{\Omega}$ is less than $10^{-9}$.
To evaluate the BCH series up to given truncated order we use the following identity~\cite{evangelista2011alternative} to express a single term with $k$ nested commutators:
\begin{equation}
	\underbrace{[[\ldots[[\hat{H},\hat{A}],\hat{A}], \ldots], \hat{A}]}_{k\text{ nested commutators}} = \sum_{l = 0}^{k}(-1)^{l} {k\choose l} \hat{A}^l\hat{H}\hat{A}^{k-l}.
\end{equation}

Our implementation of the ic-MRUCC method closely follows the algorithm used in the ic-MRCC method.\cite{evangelista2011orbital} To solve the ic-MRUCC amplitude equations, an approximated quasi-Newton method is employed.
Following Ref.~\citenum{hanauer2012communication}, instead of working with GNO operators directly, we use a simpler strategy based on the transformation matrix $\mathbf{G}$ which connects the normal-ordered operators ($\hat{\mathbf{O}}_{\mathrm{GNO}}$) to the corresponding bare operators ($\hat{\mathbf{O}}$):
\begin{equation}
\hat{\mathbf{O}}_{\mathrm{GNO}} =\hat{\mathbf{O}}\mathbf{G}.
\end{equation}
After solving the ic-MRUCC amplitude equations, the $\mathbf{G}$ matrix is constructed to transform all excitation operators for the EOM extension of ic-MRUCC to the GNO basis. 
The detailed derivation of the GNO transformation matrix $\mathbf{G}$ and its explicit block structure are provided in Appendix \ref{app:gmat}.

\section{Results}
\label{sec:results}
\subsection{The \ce{BeH2} model}
\label{sec:beh2}

\begin{figure}[h!]
\centering
\includegraphics[width=3.in]{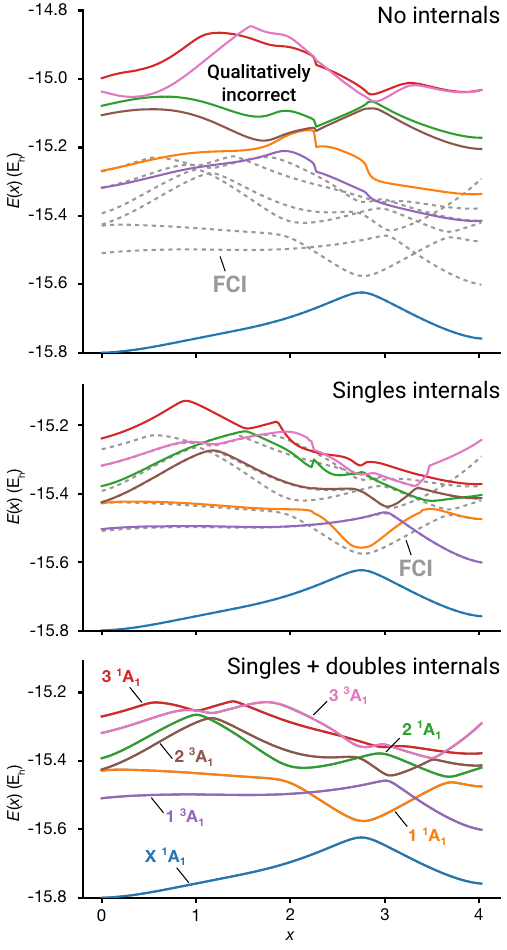} 
\caption{\label{fig:BeH2_curves}
Potential energy curves of seven low-lying states of the \ce{BeH2} model computed with the EOM-ic-MRUCCSD using different definitions of the EOM internal excitation operator.
FCI energies are indicated with dashed gray lines.
}
\end{figure}

For our first test of the EOM-ic-MRUCCSD method we consider the \ce{BeH2} model system,\cite{purvis1982full, purvis1983c2v} a well-established benchmark for new multireference methods.~\cite{evangelista2011orbital, hoffmann1988unitary, chen2012orbitally, gdanitz1988averaged, sharp2000sigma, kallay2002general, ruttink2005multireference, pittner2004performance, lyakh2006multireference}
The \ce{BeH2} model captures the salient features of the perpendicular insertion of a beryllium atom into a \ce{H2} molecule by a one-dimensional path constrained to $C_\mathrm{2v}$ symmetry.
For this model, we compute the potential energy curves (PECs) for the ground ($\mathrm{X}\ ^1\mathrm{A}_1$) and six excited states of singlet and triplet symmetry ($1\ ^1\mathrm{A}_1$, $2\ ^1\mathrm{A}_1$, $3\ ^1\mathrm{A}_1$, $1\ ^3\mathrm{A}_1$, $2\ ^3\mathrm{A}_1$, and $3\ ^3\mathrm{A}_1$).
A zeroth-order description of the ground state of this system requires two closed-shell determinants.
However, in our computations, to capture both the ground and excited states, we employ a full-valence complete active space reference state that includes the 1s orbitals of H and the 2s and 2p orbitals of Be [CASSCF(4e,6o)], and all electrons are correlated.
In the \ce{BeH2} model, the beryllium atom is placed at the center of a two-dimensional Cartesian coordinate system, and the coordinates of the two hydrogen atoms are described by the curve $y(x) = \pm(2.54 - 0.46x)$, where $x \in[0,4]\ a_0$ is the reaction coordinate.
A custom double-zeta quality basis set, \ce{Be}(10s3p/3s2p), \ce{H}(4s/2s), is used here, following earlier works.~\cite{evangelista2011orbital, chen2012orbitally}
A linear dependence threshold $\eta = 10^{-4}$ is used in the ic-MRUCC computations.

We begin by investigating how the choice of internal EOM excitations affects the accuracy of the excited states. 
In \cref{fig:BeH2_curves}, we plot the PECs for the seven electronic states of the \ce{BeH2} model system using EOM-ic-MRUCCSD with three different choices of internal EOM excitations.
When internal excitations are omitted entirely ($\hat{\mathcal{R}}^\mathrm{int} = 0$), all excited states are shifted to significantly higher energy, and the corresponding curves incorrectly predict many of the qualitative features of the FCI results.
The inclusion of single internal excitations ($\single$) introduces the necessary degrees of freedom to capture the qualitative features of several low-energy excited states ($1\ ^1\mathrm{A}_1$, $1\ ^3\mathrm{A}_1$, and $2\ ^3\mathrm{A}_1$).
However, this modification still results in qualitatively inaccurate curves for the higher-energy states ($2\ ^1\mathrm{A}_1$, $3\ ^1\mathrm{A}_1$, and $3\ ^3\mathrm{A}_1$).
Finally, in the bottom panel of \cref{fig:BeH2_curves}, we see that adding the internal doubles ($\double$) provides qualitatively correct PECs indistinguishable from the FCI ones.

\begin{figure}[htbp]
\centering
\includegraphics[width = 3.in]{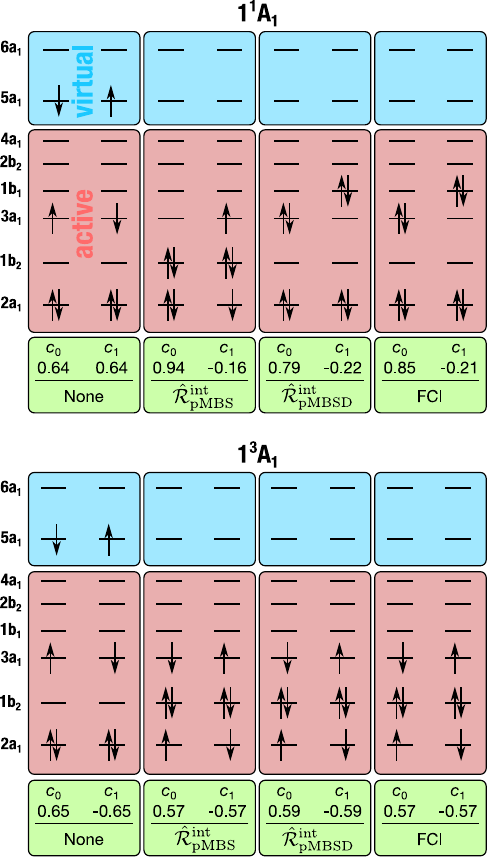} 
\caption{\label{fig:leading}
Leading determinants for the $1\ ^1\mathrm{A}_1$ and $1\ ^3\mathrm{A}_1$ excited states of the \ce{BeH2} model system (at $x$ = 2.25 $a_0$) using three different choices of internal EOM excitations and the FCI.
Orbitals in the red section represent active orbitals, while orbitals in the blue section represent virtual orbitals.
The core orbital ($1a_1$) and virtual orbitals higher than $6a_1$ are not included in the figure.
}
\end{figure}

To understand how internal excitations improve the description of excited states, we analyze the leading determinants in the EOM-ic-MRUCCSD wavefunction.
In \cref{fig:leading}, we plot the leading determinants for two representative excited states ($1\ ^1\mathrm{A}_1$ and $1\ ^3\mathrm{A}_1$) of the \ce{BeH2} model system (at $x = 2.250\ a_0$) using the three different choices of EOM internal excitations. 
The EOM-ic-MRUCCSD calculations start from the same $\mathrm{X}\ ^1\mathrm{A}_1$ reference state ($\Phi_0$), where the two leading determinants are as follows:
\begin{equation}
\begin{split}
	\ket{\Phi_0} = \phantom{-} 0.9672 &\ket{(1a_1)^2(2a_1)^2(1b_2)^2} \\
	- 0.1206 &\ket{(1a_1)^2(2a_1)^2(3a_1)^2} + \ldots,
\end{split}
\end{equation}
which differs slightly from the FCI ground state (with coefficients for the two leading determinants being $0.9652$ and $-0.1204$).
The FCI results in \cref{fig:leading} show that both the $1\ ^1\mathrm{A}_1$ and $1\ ^3\mathrm{A}_1$ states are connected to the ground state via internal excitations.
Consequently, omitting internal excitations entirely excludes important determinants and leads to the qualitatively incorrect curves shown in \cref{fig:BeH2_curves}.
The inclusion of internal single excitations ($\single$) successfully generates the correct leading determinants for the $1\ ^3\mathrm{A}_1$ state but leads to an incorrect zeroth-order description for the $1\ ^1\mathrm{A}_1$ state, which is dominated by doubly excited determinants with respect to the leading reference determinant $\ket{(1a_1)^2(2a_1)^2(1b_2)^2}$.
Finally, adding the internal double excitations ($\double$) results in the correct leading determinants for the $1\ ^1\mathrm{A}_1$ state, explaining the superior accuracy of the PECs computed at this level.

\begin{figure}[htbp]
\centering
\includegraphics[width = 3.375in]{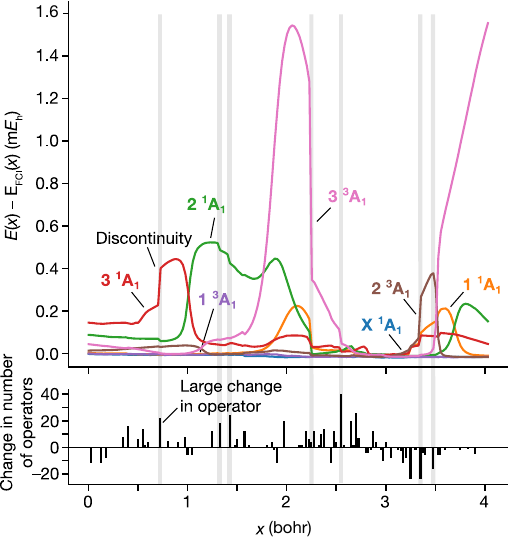}
\caption{\label{fig:BeH2_analysis}
Top panel: potential energy error curves (with respect to FCI) of seven low-lying states of the \ce{BeH2} model computed with the EOM-ic-MRUCCSD method and internal singles and doubles.
Lower panel: the change in the number of orthogonal operators that enter $\hat{\mathcal{R}}$ with respect to the previous point (to the left).
}
\end{figure}

It is important to note that all energy curves from EOM-ic-MRUCCSD are discontinuous. 
This is a common challenge encountered in multireference theories, including the MR-EOMCC method~\cite{datta2012multireference} and ic-MRCC approaches.~\cite{evangelista2011orbital}
This problem is due to the change in the number of linearly independent operators that enter the $\hat{T}$ and $\hat{\mathcal{R}}$ operators across the potential energy surface.~\cite{evangelista2011orbital}
In the case of \ce{BeH2}, this effect is more pronounced in the truncation schemes that omit internal excitations or include only internal singles, particularly around $x = 2.250$ $a_0$.
In the PECs computed with up to internal doubles, the discontinuities are less pronounced.
To analyze the origin of these discontinuities, in \cref{fig:BeH2_analysis} we plot the change in energy error and number of orthogonal operators for the $2\ ^1\mathrm{A}_1$ and $3\ ^1\mathrm{A}_1$ states.
This plot shows that discontinuities are correlated with changes in the number of linearly independent operators in $\hat{\mathcal{R}}$.

\begin{figure}[htbp]
\centering
\includegraphics[width = 3.375in]{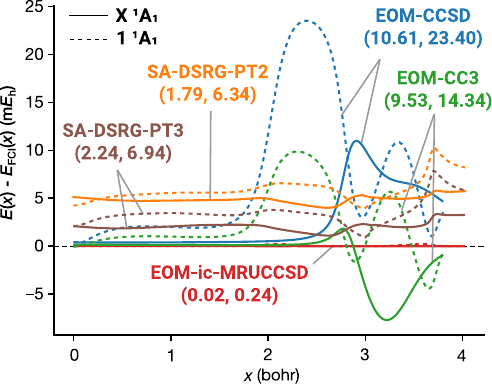} 
\caption{\label{fig:BeH2_comparison}
Potential energy error curves of different methods with respect to FCI for $\mathrm{X}\ ^1\mathrm{A}_1$ and $1\ ^1\mathrm{A}_1$ states of the \ce{BeH2} model system.
For each method the nonparallelity errors (in m$E_\mathrm{h}$) are shown in parentheses ($\mathrm{X}\ ^1\mathrm{A}_1$, $1\ ^1\mathrm{A}_1$).
}
\end{figure}

In \cref{fig:BeH2_comparison}, we compare the accuracy of EOM-ic-MRUCCSD with state-of-the-art single- and multi-reference excited-state methods, including EOM-CCSD,\cite{stanton1993equation} EOM-CC3,\cite{christiansen.1995.10.1063/1.470315} state-averaged MR-DSRG second- and third-order perturbation theory (SA-DSRG-PT2/3),\cite{li2015multireference,li2017driven,li2018driven} by plotting energy differences from the FCI value for the $\mathrm{X}\ ^1\mathrm{A}_1$ and $1\ ^1\mathrm{A}_1$ states of the \ce{BeH2} system.
The EOM-CCSD and EOM-CC3 calculations are carried out with the \textsc{Psi4} package, and the SA-DSRG-PT2/3 calculations are carried out with the \textsc{Forte} package.
To quantify the accuracy of these methods, each curve in \cref{fig:BeH2_comparison} is accompanied by the corresponding nonparallelity error (NPE), defined over a range of geometries ($X$) as:
\begin{equation}
	\mathrm{NPE} = \max_{x\in X}[\Delta E(x)] - \min_{x\in X}[\Delta E(x)],
\end{equation}
where $\Delta E(x)$ is the error with respect to the FCI energy.
Note that for EOM-CCSD and EOM-CC3, we use the following Slater determinant as the zeroth-order wavefunction:
\begin{equation}
	\ket{\Phi_1} = \ket{(1a_1)^2(2a_1)^2(1b_2)^2}.
\end{equation}
However, this is the dominant determinant for the ground state only for those geometries with $x < 2$ $a_0$.
Consequently, the \ce{BeH2} model system is particularly challenging for these single-reference methods.
Notably, both $\mathrm{X}\ ^1\mathrm{A}_1$ and $1\ ^1\mathrm{A}_1$ PECs for EOM-CCSD and EOM-CC3 exhibit significant errors within the strongly correlated region ($x > 2.825$ $a_0$),~\cite{evangelista2011orbital} yield large NPEs, and fail when $x > 3.800$ $a_0$ due to a lack of convergence of the ground state computations.
The SA-DSRG-PT2/3 methods directly target ground- and excited-state solutions by employing a state-averaged description of dynamical electron correlation.
These methods display errors consistent across the potential energy curve, including the multireference region, and yield NPEs smaller than the single-reference methods.
In comparison, the EOM-ic-MRUCCSD method stands out as the most accurate, yielding the smallest NPEs (0.02 and 0.24 m$E_h$ for $\mathrm{X}\ ^1\mathrm{A}_1$ and $1\ ^1\mathrm{A}_1$ states, respectively) among all methods.

\subsection{The dissociation curve of HF}
Our second benchmark considers the ground ($\mathrm{X}\ ^1\mathrm{A}_1$) and excited ($1\ ^1\mathrm{A}_1$, $1\ ^3\mathrm{A}_1$, and $2\ ^3\mathrm{A}_1$) PECs of hydrogen fluoride.~\cite{das2010full, evangelista2011orbital}
A consistent zeroth-order description of the ground state of this system requires considering the $\sigma$ H-F bonding and antibonding orbitals.
However, similar to the \ce{BeH2} model system, we employ a larger active space that includes the 1s orbital of \ce{H} and the 2s, 2p, and 3s orbitals of \ce{F} [CASSCF(8e,6o)].
In all computations, the fluorine 1s-like molecular orbital is excluded from the correlated computations, and the linear dependence threshold $\eta$ is set to $10^{-4}$.
We employ the DZV basis set adopted in Ref.~\citenum{das2010full}.

\begin{figure}[htbp]
\centering
\includegraphics[width = 3.375in]{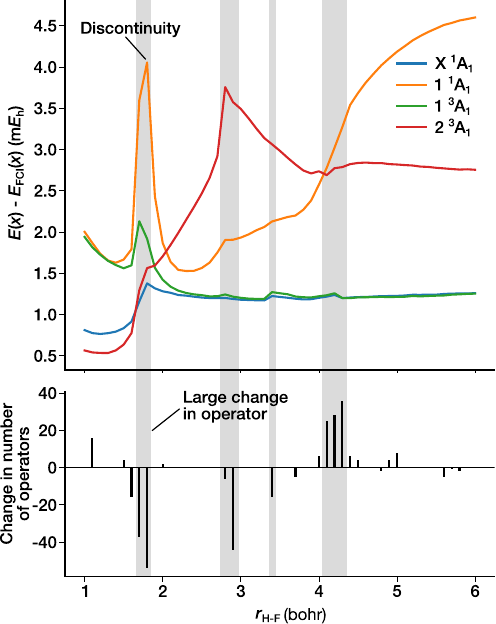} 
\caption{\label{fig:hf}
Top panel: potential energy error curves (with respect to FCI) of four low-lying states of the \ce{HF} system computed with the EOM-ic-MRUCCSD method and internal singles and doubles. 
Lower panel: the change in the number of orthogonal operators that enter $\hat{\mathcal{R}}$ with respect to the previous point (to the left).
}
\end{figure}

As shown in Fig.~S1 of the Supplementary Information, EOM-ic-MRUCCSD yields qualitatively correct PECs for the \ce{HF} molecule.
Deviations of the EOM-ic-MRUCCSD PECs from FCI are small for the ground state ($<$ 1.4 m$E_\mathrm{h}$), while they are larger for the excited states, with maximum errors of 4.6, 2.1, and 3.8 m$E_\mathrm{h}$ for the $1\ ^1\mathrm{A}_1$, $1\ ^3\mathrm{A}_1$, and $2\ ^3\mathrm{A}_1$ states, respectively (in all cases less than 0.15 eV).
Similar to the \ce{BeH2} system, all PECs are discontinuous. In \cref{fig:hf}, we plot the potential energy error curves for four states and track how the number of orthogonal operators entering $\hat{\mathcal{R}}$ changes. These discontinuities align with shifts in the number of linearly independent operators, consistent with our findings for the \ce{BeH2} model.

\begin{figure*}[t]
\centering
\includegraphics[width = 6.25in]{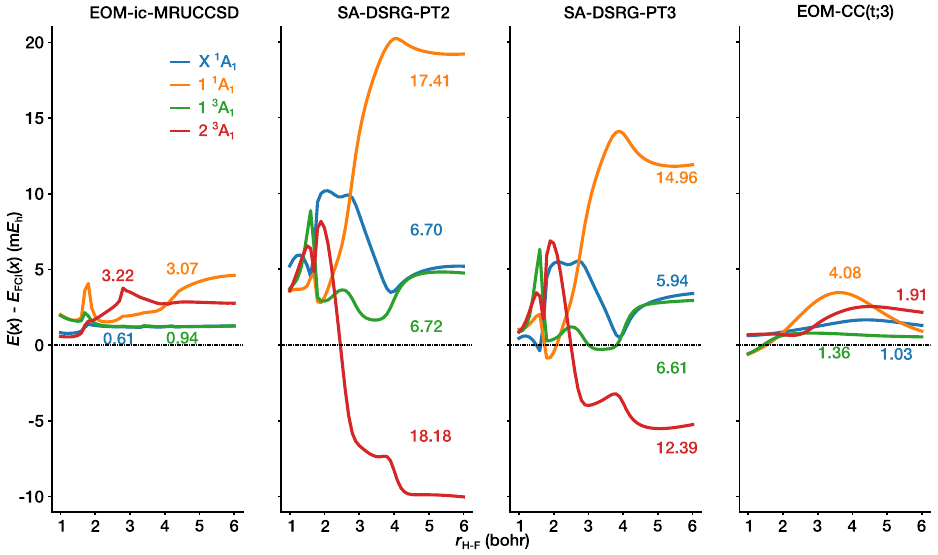} 
\caption{\label{fig:hf_compare}
Potential energy error curves of different methods with respect to FCI for $\mathrm{X}\ ^1\mathrm{A}_1$,  $1\ ^1\mathrm{A}_1$, $1\ ^3\mathrm{A}_1$, and $2 ^3\mathrm{A}_1$ states of the \ce{HF} molecule.
For each method, the nonparallelity errors (in m$E_\mathrm{h}$) are shown alongside the curves. 
}
\end{figure*}

We also compare the EOM-ic-MRUCCSD method with other excited-state methods.
The potential energy error curves for different states obtained with EOM-ic-MRUCCSD, SA-DSRG-PT2, SA-DSRG-PT3, and EOM-CC$(\mathrm{t};3)$\cite{shen.2012.10.1063/1.3700802} are shown in \cref{fig:hf_compare}.
We also computed EOM-CCSDt\cite{kowalski.2000.10.1063/1.1318757} curves, but these are not shown as they are nearly indistinguishable from the EOM-CC$(\mathrm{t};3)$ curves.
The EOM-CCSDt and EOM-CC$(\mathrm{t};3)$ calculations are carried out in the \textsc{CCpy} package,\cite{ccpy} which is a plug-in to the \textsc{PySCF} package.\cite{sun.2020.10.1063/5.0006074}
EOM-CCSD and EOM-CC3 computations did not yield continuous PECs at stretched geometries and, therefore, are not included in this figure.
The EOM-ic-MRUCCSD curves exhibit minimal errors and have the lowest NPE values among all methods except for the $2\ ^3\mathrm{A}_1$ state.
For this single-bond breaking problem, the EOM-CC$(\mathrm{t};3)$ curves are smooth and yield comparable NPE values.

\subsection{Symmetric dissociation of the water molecule}
Our next application focuses on the symmetric dissociation of the water molecule.
This system has been utilized as a benchmark for various theories and serves as a prototype for multiple bond-breaking processes.~\cite{evangelista2011orbital, yanai2006canonical, kallay2002general, szalay1995approximately, bauschlicher1986benchmark, olsen1996full, li2003pair, mahapatra2008molecular,bauman.2017.10.1080/00268976.2017.1350291}
Following Ref.~\citenum{evangelista2011orbital}, we calculate PECs for the two lowest singlet states ($\mathrm{X}\ ^1\mathrm{A}_1$ and $1\ ^1\mathrm{A}_1$) and two lowest triplet excited states ($1\ ^3\mathrm{A}_1$ and $2\ ^3\mathrm{A}_1$) for the symmetric dissociation path with the \ce{H-O-H} bond angle constrained to 109.57\degree, while scanning the \ce{O-H} bond distance within the range of $[r_e, 4r_e]$ at $0.1$ $r_e$ spacing, where $r_e = 0.9929$ \AA.
We employ the 6-31G basis and select an active space that includes the 1s orbital of hydrogen and the 2p manifold for the oxygen atom, resulting in a CASSCF(6e,5o) reference wavefunction.
The oxygen 1s-like molecular orbital is excluded from correlated computations.
The linear dependence threshold $\eta$ is set to $10^{-5}$ for the ground ($\mathrm{X}\ ^1\mathrm{A}_1$) and two triplet excited states ($1\ ^3\mathrm{A}_1$ and $2\ ^3\mathrm{A}_1$).
For the singlet excited state, we employ two thresholds ($\eta_{1}$ and $\eta_{2}$).
The $\eta_{1}$ threshold ($10^{-5}$) is used to eliminate linear dependencies in the singles, semi-internal doubles (with at least one active orbital creation and annihilator), and internal excitations.
The remaining doubles typically exhibit less severe linear dependencies, and we employ a smaller truncation threshold $\eta_{2}$ ($10^{-8}$).
Employing two different thresholds is essential for obtaining a qualitatively correct PEC for the $1\ ^1\mathrm{A}_1$ excited state. We will examine this aspect after analyzing the results obtained with two thresholds.

\begin{figure}[htbp]
\centering
\includegraphics[width = 3.375in]{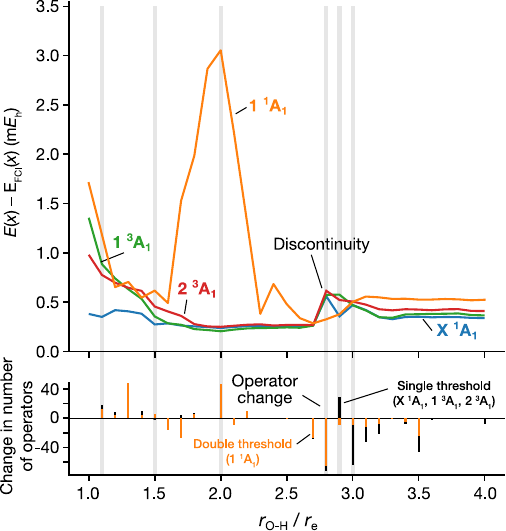} 
\caption{\label{fig:water}
Top panel: potential energy error curves (with respect to FCI) for the $\mathrm{X}\ ^1\mathrm{A}_1$, $1\ ^1\mathrm{A}_1$, $1\ ^3\mathrm{A}_1$, and $2\ ^3\mathrm{A}_1$ states of water along the symmetric dissociation path computed with the EOM-ic-MRUCCSD method and internal projected many-body singles and doubles.
Two thresholds $\eta_{1} = 10^{-5}$ and $\eta_{2} = 10^{-8}$ are used in the EOM step to eliminate linear dependencies for the $1\ ^1\mathrm{A}_1$ state, while all other curves used a single orthogonalization threshold $\eta = 10^{-5}$.
Lower panel: the change in the number of orthogonal operators and the energy errors with respect to the previous point (to the left).
}
\end{figure}

In \cref{fig:water}, we plot the potential energy error curves for all states along with the change in the number of orthogonal operators that enter into $\hat{\mathcal{R}}$.
Throughout the sampled range of geometries, the EOM-ic-MRUCCSD PECs for the $\mathrm{X}\ ^1\mathrm{A}_1$, $1\ ^3\mathrm{A}_1$, and $2\ ^3\mathrm{A}_1$ states show very small deviations from FCI (less than 1.4 m$E_{\mathrm{h}}$, within chemical accuracy).
In contrast, the $1\ ^1\mathrm{A}_1$ state shows errors as large as ca. 3 m$E_{\mathrm{h}}$.
Consistent with our previous findings, all PECs exhibit small discontinuities that correlate with changes in the number of linearly independent EOM operators.

\begin{figure}[htbp]
\centering
\includegraphics[width = 3.375in]{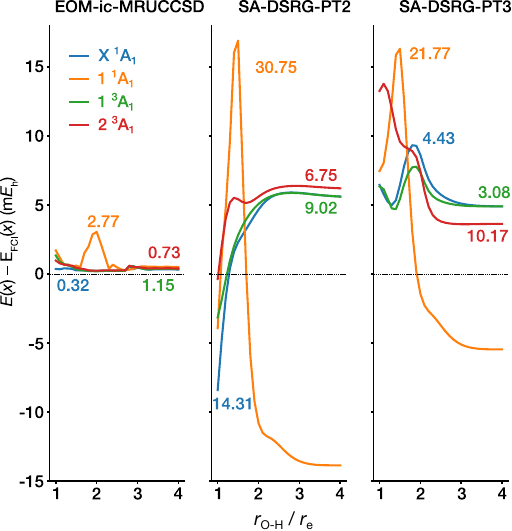} 
\caption{\label{fig:water compare}
Potential energy error curves of different methods with respect to FCI for $\mathrm{X}\ ^1\mathrm{A}_1$, $1\  ^1\mathrm{A}_1$, $1\ ^3\mathrm{A}_1$, and $2\ ^3\mathrm{A}_1$ states of the symmetric dissociation of the water.
For each method, the nonparallelity errors (in m$E_\mathrm{h}$) are shown alongside the curves. 
}
\end{figure}

We also compared the EOM-ic-MRUCCSD method with the SA-DSRG-PT2 and SA-DSRG-PT3 methods, while we excluded EOM-CCSD and EOM-CC3 since they cannot be consistently converged along the dissociation path.
As shown in \cref{fig:water compare}, the EOM-ic-MRUCCSD curves yield the smallest NPEs (0.32, 2.77, 1.15, and 0.73 m$E_\mathrm{h}$ for the $\mathrm{X}\ ^1\mathrm{A}_1$, $1\ ^1\mathrm{A}_1$, $1\ ^3\mathrm{A}_1$, and $2\ ^3\mathrm{A}_1$ states, respectively) among all methods. The PECs obtained from SA-DSRG-PT3 generally have smaller NPEs compared to SA-DSRG-PT2, except for the $2\ ^3\mathrm{A}_1$ state (6.75 and 10.17 m$E_{\mathrm{h}}$ for SA-DSRG-PT2 and PT3, respectively).
However, errors from SA-DSRG methods consistently exceed those from EOM-ic-MRUCCSD.

In contrast to the HF single bond-breaking test case, where variants of single-reference EOM-CC could accurately describe all curves, these methods encounter difficulties describing the ground and excited states of water along the double dissociation path.
This is expected as previous reports show that these methods are better suited for single bond-breaking processes,\cite{gururangan.2025.10.1016/j.cplett.2024.141840} and quadruple excitations are necessary for the accurate description of double dissociation processes.\cite{bauman.2017.10.1080/00268976.2017.1350291}
In Fig.~S4 of the Supplementary Information, we show the PECs of these states computed with EOM-CCSDt, EOM-CC(t;3), EOM-CCSDT, and CCSDTQ (only the ground state).
The EOM methods exhibit qualitatively incorrect PECs for the ground and singlet excited states, and all ground-state computations fail to converge for $r$(O--H)$>3.0\ r_e$.

Next, we discuss the challenges encountered in the computations on the $1\ ^1\mathrm{A}_1$ state.
As shown in Fig.~S2 of the Supplementary Information, EOM-ic-MRUCCSD produces a qualitatively incorrect PEC when using a single orthogonalization threshold in the EOM step ($\eta = 10^{-5}$).
A notable discontinuity appears in the dissociation region around $r (\ce{O-H}) = 3.5\ r_e$ for the single-threshold calculation, while this discontinuity is absent in the double-threshold calculation.
As shown in Fig.~S3 of the Supplementary Information, the pronounced discontinuity in the single-threshold calculation is correlated with the change in the number of linearly dependent operators that enter into $\hat{\mathcal{R}}$. 
The energy error significantly increases at $r (\ce{O-H}) = 3.5\ r_e$ as 46 operators are removed from the EOM operator space, resulting in a notable discontinuity in the PEC.

To understand the impact of these removed operators on the excited-state energy in the single-threshold calculation, we reintroduce them into the operator space and measure the resulting energy corrections for the $1\ ^1\mathrm{A}_1$ state. 
We find that operators contributing significant energy corrections (larger than $0.1\ \mathrm{m}E_{\mathrm{h}}$) are linear combinations of the form $\hat{a}_{u_{\alpha} u_{\beta}}^{e_\alpha f_\beta}$  where $u$ is one of the active orbitals $(1b_{2}, 2b_{2}, 3a_1, 4a_1)$, and $e$ and $f$ are virtual orbitals.
These operators replace an electron pair from a single active orbital with two virtual spin orbitals.
The reason why in the dissociated region, these operators are excluded from the EOM ansatz of the PEC is that the $\mathrm{X}\ ^1\mathrm{A}_1$ ic-MRUCCSD state becomes largely dominated by open-shell determinants with singly occupied $1b_2$, $2b_2$, $3a_1$, and $4a_1$ orbitals.
For example, at  $r (\ce{O-H}) = 3.5\ r_e$, the ground state is given by
\begin{align}
\ket{\Psi_0}= & \ 0.5464 \, [\ket{(2a_1)^2 (1b_1)^2 1b_{2\alpha} 3a_{1\beta} 2b_{2\beta} 4a_{1\alpha}} \notag \\
&+ \ket{(2a_1)^2 (1b_1)^2 1b_{2\beta} 3a_{1\alpha} 2b_{2\alpha} 4a_{1\beta}}] + \cdots.
\end{align}
Determinants with paired active electrons contribute only minorly to $\ket{\Psi_0}$, and so applying $\hat{a}_{u_{\alpha} u_{\beta}}^{e_\alpha f_\beta}$ to this state generates a term with near zero norm.
Consequently, these operators are removed during the orthogonalization step, causing a cumulative error on the order of 5 m$E_\mathrm{h}$.
This issue arises because the operators are orthogonalized using a metric derived solely from the ground-state wavefunction; as a result, they can be discarded simply because they (approximately) annihilate the ground state, even though they are important for describing certain excited states.  
We note that while this problem may be solved using two sets of thresholds, yielding a qualitatively correct PEC in the range $r(\ce{O-H}) \in [r_e, 4r_e]$, this discontinuity is expected to reappear at a larger bond distance, where the eigenvalues of pair excitation operators approach zero.
Indeed, at $r(\ce{O-H}) = 6\ r_e$, and even setting $\eta_{2} = 10^{-10}$ fails to capture all important pair excitation operators, resulting in a similar 6 m$E_{\mathrm{h}}$ energy error in the double-threshold PEC.
It is important to note that since this issue is independent of the specific treatment of correlation (the orthogonalization is based on the reference state only), it is expected to affect \emph{all} EOM-like multireference methods based on internally contracted ans\"{a}tze.


\subsection{Truncation of the commutator series}
Since the BCH expansion of the unitarily transformed Hamiltonian does not terminate, any practical implementation of EOM-ic-MRUCC must evaluate this operator in an approximate way.
One approach is to approximate $\bar{H}$ with a BCH expansion truncated at a given order.
To evaluate the impact of this approximation, we perform a statistical analysis of the error introduced in the ground- and excited-state PECs of the \ce{BeH2}, \ce{HF}, and water systems.

The BCH expansion is truncated to the same order in all expressions involving $\bar{H}$, including the ic-MRUCC energy and amplitude equations and the EOM-ic-MRUCC eigenvalue equation.
\cref{table:truncate BCH} shows the root-mean-square error (RMSE) for the energy of all states computed with a truncated $\bar{H}$ with respect to the non-truncated EOM-ic-MRUCCSD results. 
The inclusion of the second commutator yields satisfactory results for the \ce{BeH2} system, with all RMSEs approximately $10^{-5}$ $E_{\mathrm{h}}$.
However, this level of truncation is inaccurate for \ce{HF} and water, as the RMSEs become larger than 1.0 m$E_{\mathrm{h}}$.
The RMSE is exceptionally large for the excited states of water, with values of 0.415, 0.486, and 0.436 $E_{\mathrm{h}}$ for $1\ ^1\mathrm{A}_1$, $1\ ^3\mathrm{A}_1$, and $2\ ^3\mathrm{A}_1$ states, respectively. 

The introduction of the triple commutator does not significantly enhance results for the \ce{BeH2} and \ce{HF} systems but substantially improves the accuracy of the water system PECs. 
In this case, all RMSEs for the water system are reduced to around $10^{-4}$ $E_{\mathrm{h}}$. 
Notably, the inclusion of the four-nested commutator leads to a significant reduction in all RMSEs, bringing them to around $10^{-6}$ $E_{\mathrm{h}}$ for the \ce{BeH2} model and $10^{-5}$ $E_{\mathrm{h}}$ for the \ce{HF} and water systems.
These results are promising and indicate that EOM-ic-MRUCC schemes with a truncated BCH expansion may serve as robust and reliable approximations to the full EOM-ic-MRUCC.
The truncation of the BCH expansion preserves the scaling and orbital invariance properties of EOM-ic-MRUCC.

\begin{table}
    \begin{center}
	\caption{Statistical analysis of the error introduced in the EOM-ic-MRUCCSD results ground- and excited-state energies when truncating the BCH expansion of the unitarily-transformed Hamiltonian. Root-mean-square error (RMSE, in $E_{\mathrm{h}}$) computed for the potential energy curves of \ce{BeH2}, \ce{HF}, and \ce{H2O}. The RMSE is computed separately for each electronic state using $N$ equally spaced samples of the potential energy curve as
$\mathrm{RMSE} = \sqrt{
{N}^{-1}
\sum\limits_{i=1}^{N} (\delta E^{(k)}_i)^{2}
}$, where $\delta E^{(k)}_i$ is the energy error with respect to the non-truncated theory for the $i$-th point on the curve when $\bar{H}$ is evaluated with up to $k$ nested commutators.}
	\label{table:truncate BCH}
        \bgroup
        \def\arraystretch{1.}
	\begin{tabular}{cccc}
		\hline
		\hline
		&\multicolumn{3}{c}{\ce{BeH2}} \\ \hline
		States & 2  &  3 & 4 \\ \hline
		$\mathrm{X}\ ^1\mathrm{A}_1$   & $4.567 \times 10^{-5}$ & $1.944 \times 10^{-5}$& $7.819 \times 10^{-6}$\\
		$1\ ^1\mathrm{A}_1$ & $1.286 \times 10^{-5}$ & $2.235 \times 10^{-5}$& $8.631 \times 10^{-6}$\\
		$2\ ^1\mathrm{A}_1$ & $1.328 \times 10^{-5}$ & $2.120\times 10^{-5}$& $9.213 \times 10^{-6}$\\ 
		$3\ ^1\mathrm{A}_1$ & $1.455 \times 10^{-5}$ & $2.014 \times 10^{-5}$& $9.713 \times 10^{-6}$\\
		$1\ ^3\mathrm{A}_1$ & $1.860 \times 10^{-5}$ & $1.862 \times 10^{-5}$& $7.102 \times 10^{-6}$\\
		$2\ ^3\mathrm{A}_1$ & $1.804 \times 10^{-5}$ & $1.684 \times 10^{-5}$& $7.142 \times 10^{-6}$\\
		$3\ ^3\mathrm{A}_1$ & $1.882 \times 10^{-5}$ & $1.643 \times 10^{-5}$& $7.937 \times 10^{-6}$\\
            \hhline{====} 
        &\multicolumn{3}{c}{\ce{HF}} \\ \hline   
        States & 2  &  3 & 4 \\ \hline
        $\mathrm{X}\ ^1\mathrm{A}_1$   & $2.928 \times 10^{-4}$ & $1.840 \times 10^{-4}$& $1.633 \times 10^{-5}$\\
        $1\ ^1\mathrm{A}_1$   & $5.121 \times 10^{-3}$ & $2.234 \times 10^{-3}$& $9.514 \times 10^{-5}$\\
        $1\ ^3\mathrm{A}_1$ & $8.798 \times 10^{-4}$ & $2.712 \times 10^{-4}$& $1.608 \times 10^{-5}$\\
		$2\ ^3\mathrm{A}_1$ & $2.677 \times 10^{-3}$ & $1.437\times 10^{-3}$& $3.640 \times 10^{-5}$\\ 

            \hhline{====}
		&\multicolumn{3}{c}{\ce{H2O}} \\ \hline
		States & 2  &  3 & 4 \\ \hline
		$\mathrm{X}\ ^1\mathrm{A}_1$   & $3.957 \times 10^{-4}$ & $2.518 \times 10^{-4}$& $1.895 \times 10^{-5}$\\
		$1\ ^1\mathrm{A}_1$   & $4.153 \times 10^{-1}$ & $2.309 \times 10^{-4}$& $2.330 \times 10^{-5}$\\
		$1\ ^3\mathrm{A}_1$ & $4.859 \times 10^{-1}$ & $1.958 \times 10^{-4}$& $1.316 \times 10^{-5}$\\
		$2\ ^3\mathrm{A}_1$ & $4.363 \times 10^{-1}$ & $1.884\times 10^{-4}$& $1.711 \times 10^{-5}$\\ 
		\hline
		\hline
	\end{tabular}
        \egroup
	\end{center}
\end{table}

\section{Conclusion}
\label{sec:conclusion}
We have reported a new orbital invariant and size-intensive equation-of-motion extension of the internally contracted multireference unitary coupled-cluster (EOM-ic-MRUCC) method.
In this approach, the underlying ground state is computed at the ic-MRUCC level, using a projective approach based on generalized normal-ordered operators, which, unlike previous formulations,\cite{chen2012orbitally} is rigorously size consistent and size extensive.
The EOM-ic-MRUCC method follows the transform-then-diagonalize route, like its non-unitary counterpart (MR-EOMCC).~\cite{datta2012multireference}
In formulating the EOM extension of ic-MRUCC, we adopt the self-consistent operator approach and analyze different operator choices.
We propose using projected many-body operators to represent the internal components of the excitation operator that modify the occupation of active orbitals.
This choice ensures polynomial scaling in the number of active orbitals while preserving orbital invariance.
Moreover, since projected many-body operators satisfy the killer condition, the excitation energies are size intensive.

Benchmark results for the \ce{Be} $+$ \ce{H2} insertion reaction, the dissociation of \ce{HF}, and the symmetric dissociation of water are obtained using a pilot code implementation of EOM-ic-MRUCC theory based on the expansion of the corresponding equations in the full configuration-interaction basis. 
Within our benchmark results, EOM-ic-MRUCC truncated to single and double excitations (EOM-ic-MRUCCSD) yields highly accurate potential energy curves that deviate from FCI by less than 5 m$E_\mathrm{h}$ and have nonparallelity errors less than 4 m$E_\mathrm{h}$.
When applied to compute excited states over an entire potential energy curve, EOM-ic-MRUCC outperforms both single-reference methods (EOM-CCSD, EOM-CC3) and low-order perturbative approximations (SA-DSRG-PT2 and SA-DSRG-PT3) across all examined excited states.
We find that truncating the BCH expansion to low commutator orders introduces negligible errors: inclusion of up to four commutators yields RMSEs less than 0.01 m$E_{\mathrm{h}}$ for the three systems considered in this work.
Properties such as orbital invariance, size consistency, and size extensivity are preserved when the BCH expansion is truncated.

Despite its many desirable properties, we identify two major challenges with the EOM-ic-MRUCC method.
Firstly, in truncated schemes, the use of projected many-body operators introduces high-order RDMs.
Although it may be tempting to approximate higher-order RDMs with lower-order RDMs by neglecting higher-order cumulants,~\cite{li2023intruder} such approximations do not satisfy $N$-representability and can lead to variational collapse.~\cite{saitow2013multireference}
Secondly, potential energy curves obtained from EOM-ic-MRUCC are discontinuous due to abrupt changes in the number of orthogonal operators, potentially leading to qualitatively incorrect results.
Discontinuities are observed in both ground and excited state potential energy curves. 
Unfortunately, these discontinuities are expected to affect any theory that expresses excited states via excitation operators applied to an internally contracted state.
These discontinuities may result in a qualitatively incorrect potential energy surface, and can be mitigated by introducing separate orthogonalization thresholds for different classes of excitations.
Future work will focus on addressing these issues by reformulating the EOM-ic-MRUCC method using the driven similarity renormalization group framework.

\section*{Supplementary Material}
The Supplementary Material contains Figs. S1-S4, which reports potential energy curves of the \ce{HF} molecule and symmetric dissociation of water, computed with various single- and multi-reference methods.

\begin{acknowledgments}
We acknowledge Prof. Alexander Yu. Sokolov for valuable discussions on the discontinuity issue of EOM-ic-MRUCC theory and thank Prof. Piotr Piecuch, Dr. Ilias Magoulas, and Mr. Karthik Gururangan for valuable comments. This research was supported by the U.S. Department of Energy under Award DE-SC0024532.
S.L. thanks the National Science Foundation and the Molecular Sciences Software Institute for the financial support under Grant No. CHE-2136142.
\end{acknowledgments}

\section*{DATA AVAILABILITY}
All data are available upon reasonable request.
The software used to produce the data presented in this work is available in an accompanying public code repository.~\cite{gitrepo}

\appendix
\setcounter{equation}{0}
\renewcommand{\appendixname}{APPENDIX}

\section{Scaling Properties and Choice of Orthogonalization Scheme} 
\label{sec:size-extensive}

This appendix discusses the formal scaling properties of ic-MRUCC and its EOM extension,\cite{brueckner1955many, hugenholtz1957perturbation,pople1976theoretical,bartlett1981many}
focusing on the role played by the choice of the orthonormal excitation operators and the pool of operators used to define the EOM extension.
A key requirement for the size extensivity of the energy (correct asymptotic scaling of correlation energy with system size) is that the diagrammatic expansion of the energy contains only connected diagrams.
When the energy of a size-extensive method is invariant with respect to orbital rotations among sets of orbitals that preserve the structure of the reference state, size extensivity also implies additivity of the energy for separable states of non-interacting fragments (size consistency).

Previous studies have shown that the size extensivity of ic-MRCC depends on the choice of the operators that enter in $\hat{T}$ and the $\mathbf{X}$ matrix defined in \cref{eq:ktx} of the main text.\cite{evangelista2011orbital,hanauer2011pilot} 
In particular, Hanauer and K{\"o}hn\cite{hanauer2012communication} demonstrated that for a basis of generalized normal-ordered operators, either a full or sequential orthogonalization procedure ensures size extensivity.
Similar considerations apply to the unitary version of ic-MRCC, and in our work, we ensure size extensivity by employing the GNO orthogonalization scheme described in \cref{app:gmat}.
The proof of the size extensivity for the GNO-based ic-MRCC follows along the same lines as for ic-MRCC,~\cite{hanauer2012communication} differing only in the transformed Hamiltonian $\bar{H}$ expression.
Like in ic-MRCC theory, the transformed Hamiltonian is also connected in the unitary version due to the nested commutator structure of the Baker--Campbell--Hausdorff (BCH) formula:
\begin{equation}
	\bar{H} = e^{-\hat{A}} \hat{H} e^{\hat{A}} = \hat{H} + \sum_{k = 1}^{\infty} \frac{1}{k!}
\underbrace{[[\ldots[[\hat{H},\hat{A}],\hat{A}], \ldots], \hat{A}]}_{k\text{ nested commutators}}.
\end{equation}
Note that when formulated via the variational method,\cite{chen2012orbitally} ic-MRUCC theory is expected to lack size extensivity due to the appearance of disconnected terms upon taking the derivative of the truncated effective Hamiltonian with respect to the amplitudes.

\begin{figure}[!htb]
    \includegraphics[width=3.375in]{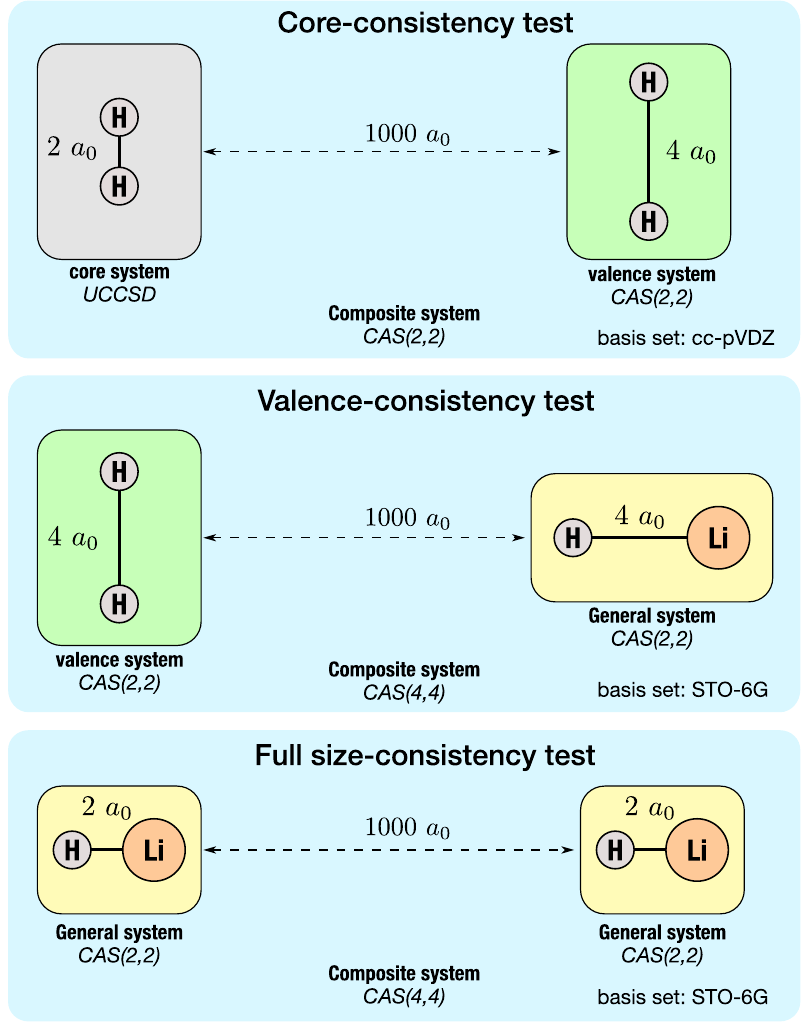}
    \caption{Systems used in testing the size-consistency properties of ic-MRUCC. A total of nine ic-MRUCCSD computations (italic labels) were performed. A ``core'' system has only core and virtual orbitals (no active orbitals), a ``valence'' system has all electrons in the active space (no core orbitals), and a ``general'' system has core, active, and virtual orbitals.}
    \label{fig:test-systems}
\end{figure}

To test size consistency numerically, we have implemented all three orthogonalization schemes considered in the literature, here denoted as ``full'',\cite{evangelista2011orbital} ``sequential'',\cite{hanauer2011pilot, kohn2020improved} and ``GNO''.\cite{hanauer2012communication}
The ``full'' and ``sequential'' schemes apply L{\"o}wdin canonical orthogonalization to the non-normal-ordered operator basis in one step or sequentially (first singles, then doubles), respectively.
The ``GNO'' scheme applies L{\"o}wdin canonical orthogonalization to GNO normal-ordered operators.
For each scheme, we have conducted three sets of numerical tests for core, valence, and full size consistency of the energy.
The systems used to test these properties are shown in \cref{fig:test-systems}.
1) The core-consistency test partitions the systems such that one \ce{H2} molecule is assigned core and virtual orbitals only, while the second \ce{H2} molecule is assigned only active and virtual orbitals.
2) The valence-consistency test considers one \ce{H2} molecule with valence and virtual orbitals and one \ce{LiH} molecule treated as a general system (with core, active, and virtual orbitals).
3) The full size-consistency test employs two \ce{LiH} molecules, both partitioned as a general system.

Table~\ref{table:size-consistency} reports the difference in energy between the composite system and the sum of its fragments. We find that only the GNO-based scheme passes all three size-consistency tests, while the full and sequential approaches fail at least one test.

\begin{table}[!htb]
\footnotesize
\caption{Core- and valence-consistency errors (in $E_\mathrm{h}$) in ic-MRUCCSD with different orthogonalization procedures. A bold entry indicates that a test has passed.}
\label{table:size-consistency}
        \bgroup
        \def\arraystretch{1}
	\begin{tabular}{lccc}
		\hline
		\hline 
		& \multicolumn{3}{c}{Size-consistency test}\\\cline{2-4}
		Orthogonalization Procedure & Core   &  Valence & Full \\ \hline
		Full	& $5.7 \times 10^{-8}$  & $2.4 \times 10^{-7}$ & $1.7 \times 10^{-4}$ \\
		Sequential & $ \bf<10^{-12}$ &  $1.5 \times 10^{-8}$ & $2.1 \times 10^{-6}$\\
		GNO &  $\bf < 10^{-12}$ & $\bf < 10^{-12}$ &  $\bf 3.2 \times 10^{-11}$\\ 
		\hline
		\hline
	\end{tabular}
        \egroup
\end{table}

\section{Proof of size intensivity for choices of the internal excitation operators}
\label{app:intensivity}
In this appendix, we prove that excitation energies computed with EOM-ic-MRUCC are size intensive.
We consider two non-interacting systems A and B, with the overall wavefunction multiplicatively separable, \textit{i.e.}, $\Psi_{\mathrm{AB}}=\Psi_{\mathrm{A}}\Psi_{\mathrm{B}}$. 
The cumulants involving orbitals localized on either A or B are identically null for such a state.
As a result, the effective Hamiltonian for the composite system $\bar{H}'_{\mathrm{AB}}$ is additively separable, \textit{i.e.}, 
$\bar{H}'_{\mathrm{AB}} = \bar{H}'_{\mathrm{A}} + \bar{H}'_{\mathrm{B}}$.

Consider an excited state of the composite system A$^*$ + B with separable wavefunction $\Psi_{\mathrm{A^*B}}=\Psi_{\mathrm{A}^*}\Psi_{\mathrm{B}}$.
Our goal is to prove that the solutions of the EOM equation for the composite system (omitting the index $\alpha$ for clarity):
\begin{equation}
	\bar{\mathbf{H}}'_{\mathrm{A+B}} \mathbf{r}_{\mathrm{A+B}} = \omega_{\mathrm{A+B}} \mathbf{S}'_{\mathrm{A+B}}\mathbf{r}_{\mathrm{A+B}},
\end{equation}
contain all the solutions of the EOM problem for the isolated system A:
\begin{equation}
	\bar{\mathbf{H}}'_{\mathrm{A}} \mathbf{r}_{\mathrm{A}} = \omega_{\mathrm{A}} \mathbf{S}'_{\mathrm{A}}\mathbf{r}_{\mathrm{A}}.
\end{equation}
To prove that EOM-ic-MRUCC satisfies this condition, it is sufficient to show that the Hamiltonian ($\bar{\mathbf{H}}'$) and metric ($\mathbf{S'}$) matrices  (in the original $\hat{\boldsymbol{\rho}}$ basis) are block-diagonal, when we partition the operators into the three sets of excitations localized on A or B and excitations involving A and B.

Since the ground state reference is assumed to be separable, the original, linearly dependent, EOM operators basis involving orbitals from both subsystems are multiplicatively separable, that is, we can write an operator as $\hat{\rho}_{\mathrm{AB}} = \tilde{\rho}_{\mathrm{A}}\tilde{\rho}_{\mathrm{B}}$, where $\tilde{\rho}_{\mathrm{A}}$ and $\tilde{\rho}_{\mathrm{B}}$ act only on A or B.
For external excitations, these operators correspond to normal-ordered products $\{ \hat{a}_{ij \cdots}^{ab\cdots}\}_X$ while for internals they are projected operators of the form $\{ \hat{a}_{uv \cdots}^{xy\cdots}\}_X \ket{\Phi_X} \bra{\Phi_X}$ where $X = A$ or $B$ denotes a restriction of the indices.
As a consequence, these operators satisfy the killer condition when applied to the individual fragment reference states.

 This property holds for the generalized normal-ordered operators since, for separable states, there are no cumulants involving orbitals from both A and B.
Elements of the effective Hamiltonian within the A-B block can be shown to be null:
\begin{align}
	& \braket{\Phi_0^{\mathrm{A}}\Phi_0^{\mathrm{B}}|
	[\hat{\rho}_{\mathrm{A}}^{\dagger}, [\bar{H}'_\mathrm{A} + \bar{H}'_\mathrm{B},\hat{\rho}_\mathrm{B}^{\vphantom{\dagger}}]]
	|\Phi_0^{\mathrm{A}} \Phi_0^{\mathrm{B}}}  \notag \\
	= & \braket{\Phi_0^{\mathrm{A}}\Phi_0^{\mathrm{B}}|
	[\hat{\rho}_{\mathrm{A}}^{\dagger}, 
	\underbrace{[\bar{H}'_\mathrm{A},\hat{\rho}_\mathrm{B}^{\vphantom{\dagger}}]}_{=0}]|\Phi_0^{\mathrm{A}} \Phi_0^{\mathrm{B}}}  \notag \\
&+ \braket{\Phi_0^{\mathrm{A}}\Phi_0^{\mathrm{B}}|
	\underbrace{[\hat{\rho}_{\mathrm{A}}^{\dagger}, [\bar{H}'_\mathrm{B},\hat{\rho}_\mathrm{B}^{\vphantom{\dagger}}]]}_{=0}|\Phi_0^{\mathrm{A}} \Phi_0^{\mathrm{B}}} 	
	\notag \\	
	= & 0,
\label{eq:AB block}
\end{align}
where we used the fact that commutators of the form $[\hat{O}_\mathrm{A},\hat{O}_\mathrm{B}]$ evaluate to zero since all contracted terms are multiplied by cumulants with indices that belong to both A and B.

Next, we consider the A/A+B block. In this case the matrix elements are also null:
\begin{align}
	& \braket{\Phi_0^{\mathrm{A}}\Phi_0^{\mathrm{B}}|
	[\hat{\rho}_{\mathrm{A}}^{\dagger}, [\bar{H}'_\mathrm{A} + \bar{H}'_\mathrm{B},\hat{\rho}_\mathrm{A + B}^{\vphantom{\dagger}}]]
	|\Phi_0^{\mathrm{A}} \Phi_0^{\mathrm{B}}}  \notag \\
	= & \braket{\Phi_0^{\mathrm{A}}\Phi_0^{\mathrm{B}}|
	[\hat{\rho}_{\mathrm{A}}^{\dagger}, 
	[\bar{H}'_\mathrm{A},\hat{\rho}_\mathrm{A + B}^{\vphantom{\dagger}}]]|\Phi_0^{\mathrm{A}} \Phi_0^{\mathrm{B}}}  \notag \\
&+ 
 \braket{\Phi_0^{\mathrm{A}}\Phi_0^{\mathrm{B}}|
	[\hat{\rho}_{\mathrm{A}}^{\dagger}, [\bar{H}'_\mathrm{B},\hat{\rho}_\mathrm{A + B}^{\vphantom{\dagger}}]]|\Phi_0^{\mathrm{A}} \Phi_0^{\mathrm{B}}} 	
	\notag \\	
	= & \braket{\Phi_0^{\mathrm{A}}\Phi_0^{\mathrm{B}}|
	\hat{\rho}_{\mathrm{A}}^{\dagger}  
	[\bar{H}'_\mathrm{A},\hat{\rho}_\mathrm{A + B}^{\vphantom{\dagger}}]|\Phi_0^{\mathrm{A}} \Phi_0^{\mathrm{B}}}  \notag \\
&+ \braket{\Phi_0^{\mathrm{A}}\Phi_0^{\mathrm{B}}|
	\hat{\rho}_{\mathrm{A}}^{\dagger}  [\bar{H}'_\mathrm{B},\hat{\rho}_\mathrm{A + B}^{\vphantom{\dagger}}]|\Phi_0^{\mathrm{A}} \Phi_0^{\mathrm{B}}} 	
	\notag \\		
	= &
\braket{\Phi_0^{\mathrm{A}}|
	\hat{\rho}_{\mathrm{A}}^{\dagger}  
	\bar{H}'_\mathrm{A} \tilde{\rho}_{\mathrm{A}} |\Phi_0^{\mathrm{A}} }  
	\underbrace{\braket{\Phi_0^{\mathrm{B}} | \tilde{\rho}_{\mathrm{B}} | \Phi_0^{\mathrm{B}}}}_{=0} \notag\\
&-
\braket{\Phi_0^{\mathrm{A}}|
	\hat{\rho}_{\mathrm{A}}^{\dagger}  
	 \tilde{\rho}_{\mathrm{A}} \bar{H}'_\mathrm{A}|\Phi_0^{\mathrm{A}} }  
	\underbrace{\braket{\Phi_0^{\mathrm{B}} | \tilde{\rho}_{\mathrm{B}} | \Phi_0^{\mathrm{B}}}}_{=0} \notag \\
&+ \braket{\Phi_0^{\mathrm{A}}|
	\hat{\rho}_{\mathrm{A}}^{\dagger}  
	 \tilde{\rho}_{\mathrm{A}} |\Phi_0^{\mathrm{A}} }  
\braket{\Phi_0^{\mathrm{B}} | \bar{H}'_\mathrm{B} \tilde{\rho}_{\mathrm{B}} | \Phi_0^{\mathrm{B}}} \notag\\
&-\braket{\Phi_0^{\mathrm{A}}|
	\hat{\rho}_{\mathrm{A}}^{\dagger}  
	 \tilde{\rho}_{\mathrm{A}} |\Phi_0^{\mathrm{A}} }  
	\underbrace{\braket{\Phi_0^{\mathrm{B}} | \tilde{\rho}_{\mathrm{B}} \bar{H}'_\mathrm{B} | \Phi_0^{\mathrm{B}}}}_{=0}.
\label{eq:AB block2}
\end{align}
In deriving this result, we have first used the killer condition (e.g., $\hat{\rho}_{\mathrm{A}}^{\dagger} \Phi_0^{\mathrm{A}} = 0$) to eliminate one of the commutators, then we decomposed the excitation operator from the A + B block as
$\hat{\rho}_{\mathrm{AB}} = \tilde{\rho}_{\mathrm{A}}\tilde{\rho}_{\mathrm{B}}$ separating terms in the expectation value into products of averages over A and B.
Three of these terms are null due to the killer condition (first, second, and fourth terms), while the third term defines a necessary and sufficient condition for the energy to be size extensive. This condition may be written as
\begin{equation}
\braket{\Phi_0^{\mathrm{A}}|
	\hat{\rho}_{\mathrm{A}}^{\dagger}  
	 \tilde{\rho}_{\mathrm{A}} |\Phi_0^{\mathrm{A}} }  
\braket{\Phi_0^{\mathrm{B}} |  \tilde{\rho}_{\mathrm{B}}^\dagger \bar{H}'_\mathrm{B} | \Phi_0^{\mathrm{B}}}^* = 0,
\end{equation}
and may be satisfied if one of the terms in the product is zero.
Since $\braket{\Phi_0^{\mathrm{A}}|
	\hat{\rho}_{\mathrm{A}}^{\dagger}  
	 \tilde{\rho}_{\mathrm{A}} |\Phi_0^{\mathrm{A}} }  $ is not generally guaranteed to be zero, the condition $\braket{\Phi_0^{\mathrm{B}} |  \tilde{\rho}_{\mathrm{B}}^\dagger \bar{H}'_\mathrm{B} | \Phi_0^{\mathrm{B}}} = 0$ ultimately determines the size-intensivity property.
This proof may be easily extended to the other blocks of the EOM Hamiltonian and the metric matrix.

\section{Detailed derivation of the generalized normal-ordering transformation matrix}
\label{app:gmat}

In this appendix, we derive the matrix $\mathbf{G}$ that transforms the operator basis into generalized normal-ordered (GNO) operators. We focus here on the case of single and double excitations, with corresponding explicit relations for the GNO operators:
\begin{align}
	\{\hat{a}^{p}_{q}\} &=  \hat{a}^{p}_{q} - \gamma^{p}_{q} \\
	\{\hat{a}^{pq}_{rs}\} &= \hat{a}^{pq}_{rs} - P(pq)P(rs)\gamma^{p}_{r}\hat{a}^{q}_{s} + P(pq)P(rs)\gamma^{p}_{r}\gamma^{q}_{s} - \gamma^{pq}_{rs},
\end{align}
where $\gamma^{p}_{q}$ and $\gamma^{pq}_{rs}$ are one- and two-body reduced density matrices (RDMs) of the reference state:
\begin{align}
	\gamma_{q}^{p} &= \braket{\Phi_0|\hat{a}_{p}^{\dagger}\hat{a}_{q}|\Phi_0}, \\
    \gamma_{rs}^{pq} &= \braket{\Phi_0|\hat{a}_{p}^{\dagger}\hat{a}_{q}^{\dagger}\hat{a}_{s}\hat{a}_{r}|\Phi_0},
\end{align}
and $P(pq)$ denotes the antisymmetric permutation operator, defined as:
\begin{equation}
	P(pq)f(p,q) = f(p,q) - f(q, p).
\end{equation}

Collecting these transformations into a matrix form naturally leads to a block matrix
$\mathbf{G}$ with the structure
\begin{equation}
\mathbf{G} = 
	\begin{pmatrix}
	    \mathbf{1} & \mathbf{G}_{01} & \mathbf{G}_{02}\\ 
		\mathbf{0} & \mathbf{1} & \mathbf{G}_{12}\\
		\mathbf{0} & \mathbf{0} & \mathbf{1} 
	\end{pmatrix},
\end{equation}
where the identity blocks $\mathbf{1}$ imply that operators within a given excitation manifold (reference, singles, or doubles) remain unchanged under the GNO transformation. The nontrivial transformations appear in the off-diagonal blocks:
\begin{align}
	(G_{12})^{i, bc}_{a, jk} &= \delta_{k}^{i}(\delta_{b}^{a}\gamma_{c}^{j} - \delta_{c}^{a}\gamma_{b}^{j}) + \delta_{j}^{i}(\delta_{c}^{a}\gamma_{b}^{k} - \delta_{b}^{a}\gamma_{c}^{k}), \\
	(G_{01})^{u}_{v} &= - \gamma_{v}^{u},  \\
	(G_{02})^{uv}_{xy} &= -2\gamma_{y}^{u}\gamma_{x}^{v} + 2\gamma_{x}^{u}\gamma_{y}^{v} - \gamma_{xy}^{uv}.
\end{align}
Here, the superscripts and subscripts on the tensor elements of $\mathbf{G}$ track the transformation from one sector (e.g., reference, denoted by 0) to another (e.g., singles or doubles). For instance, $(G_{01})^{u}_{v}$ represents the part of the transformation matrix that couples the reference to singles.

%

\clearpage
\newpage
\renewcommand{\thefigure}{S\arabic{figure}}
\setcounter{figure}{0}
\setcounter{page}{1}
\renewcommand{\thepage}{S\arabic{page}}

\renewcommand{\thesection}{S\Roman{section}}
\setcounter{section}{0}

\onecolumngrid
\fontsize{12}{24}\selectfont
\begin{center}
	\textbf{\large Supplementary Information:\\
	Equation-of-motion internally contracted multireference unitary coupled-cluster theory
	}\\[.2cm]
	Shuhang Li$^{1,*}$, Zijun Zhao$^{1}$, and Francesco A.\ Evangelista$^{1,*}$\\[.1cm]
	{\itshape ${}^1$Department of Chemistry and Cherry Emerson Center for Scientific Computation,\\ 
	Emory University, Atlanta, Georgia 30322, USA\\}
	${}^*$ E-mails: shuhang.li@emory.edu, 
	francesco.evangelista@emory.edu.
\end{center}

\newpage
\noindent

\begin{figure}[htbp]
\centering
\includegraphics[width = 6.75in]{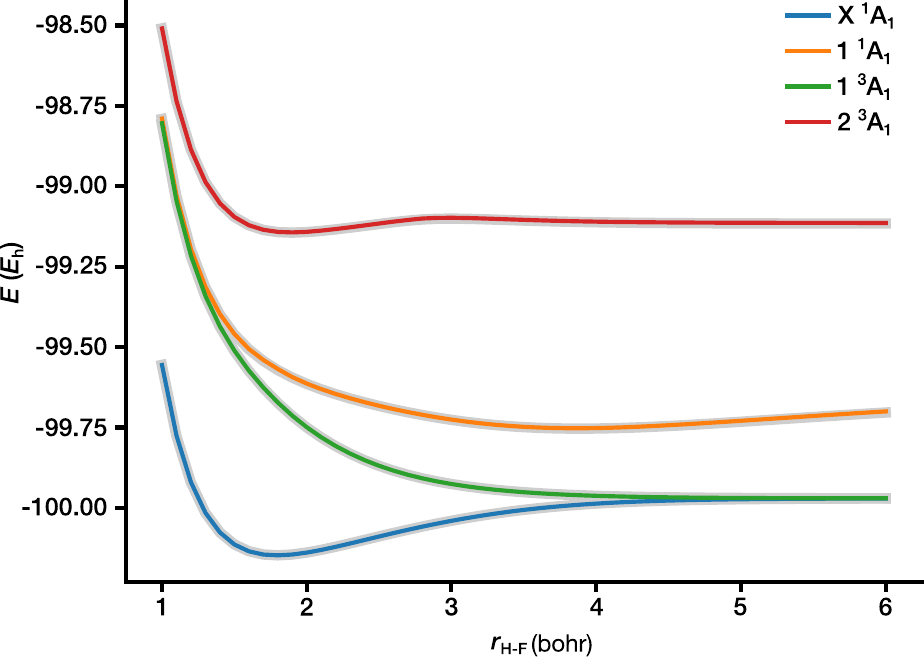} 
\caption{
Potential energy curves of FCI (gray) and EOM-ic-MRUCCSD (colored) for the $\mathrm{X}\ ^1\mathrm{A}_1$,  $1\ ^1\mathrm{A}_1$, $1\ ^3\mathrm{A}_1$, and $2 ^3\mathrm{A}_1$ states of the \ce{HF} molecule as a function of the bond distance ($r_\text{H-F}$ in \AA).
}
\end{figure}

\begin{figure}[htbp]
\centering
\includegraphics[width = 6.75in]{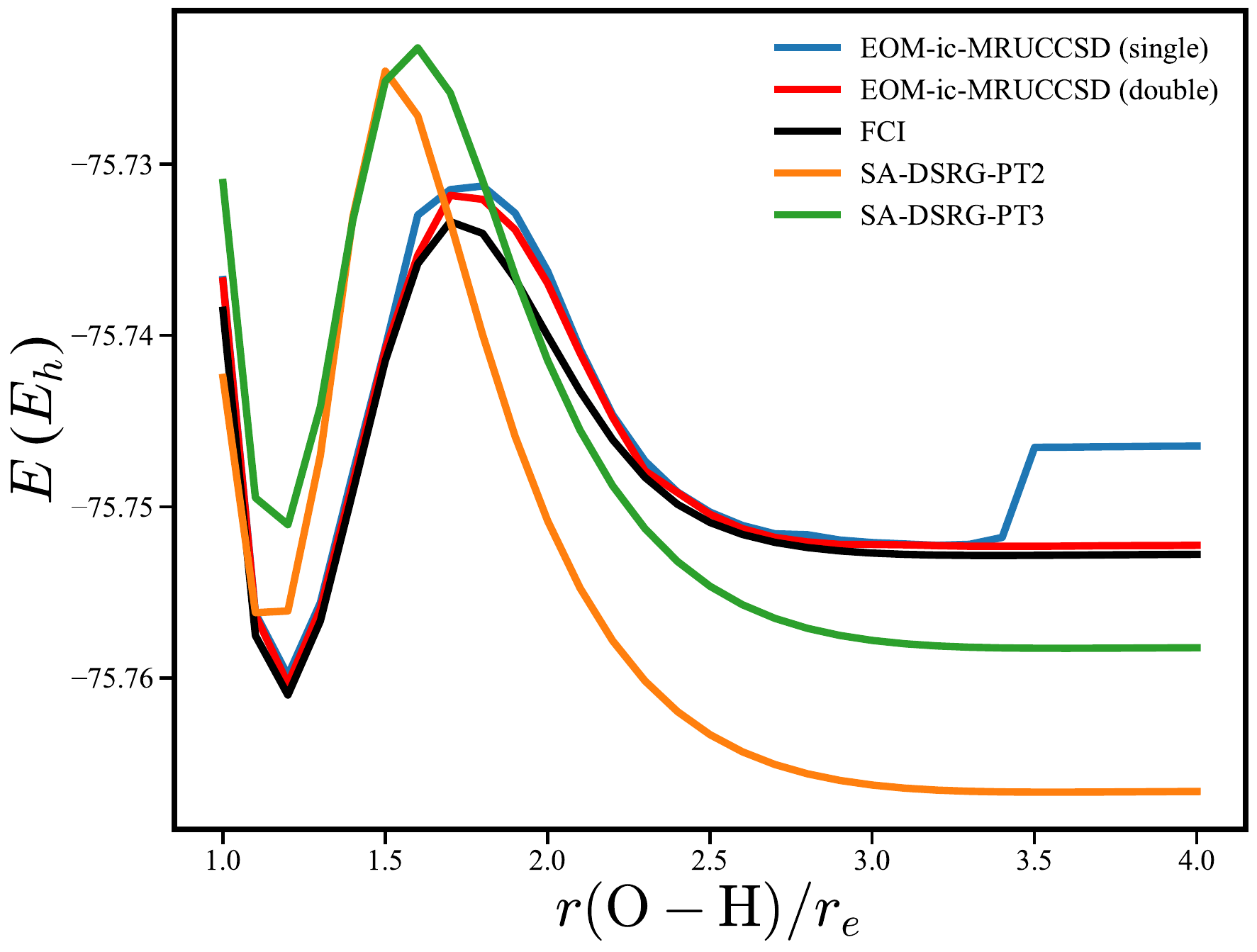} 
\caption{
Potential energy curves for the $1\ ^1\mathrm{A}_1$ state of the water symmetric dissociation path computed with different multireference methods.  
This plot highlights the different behavior of the EOM-ic-MRUCCSD method when using one orthogonalization threshold (single) $\eta = 10^{-5}$ vs. two distinct orthogonalization thresholds (double) $\eta_{1} = 10^{-5}$ and $\eta_{2} = 10^{-8}$.
}
\end{figure}

\begin{figure}[htbp]
\centering
\includegraphics[width = 6.75in]{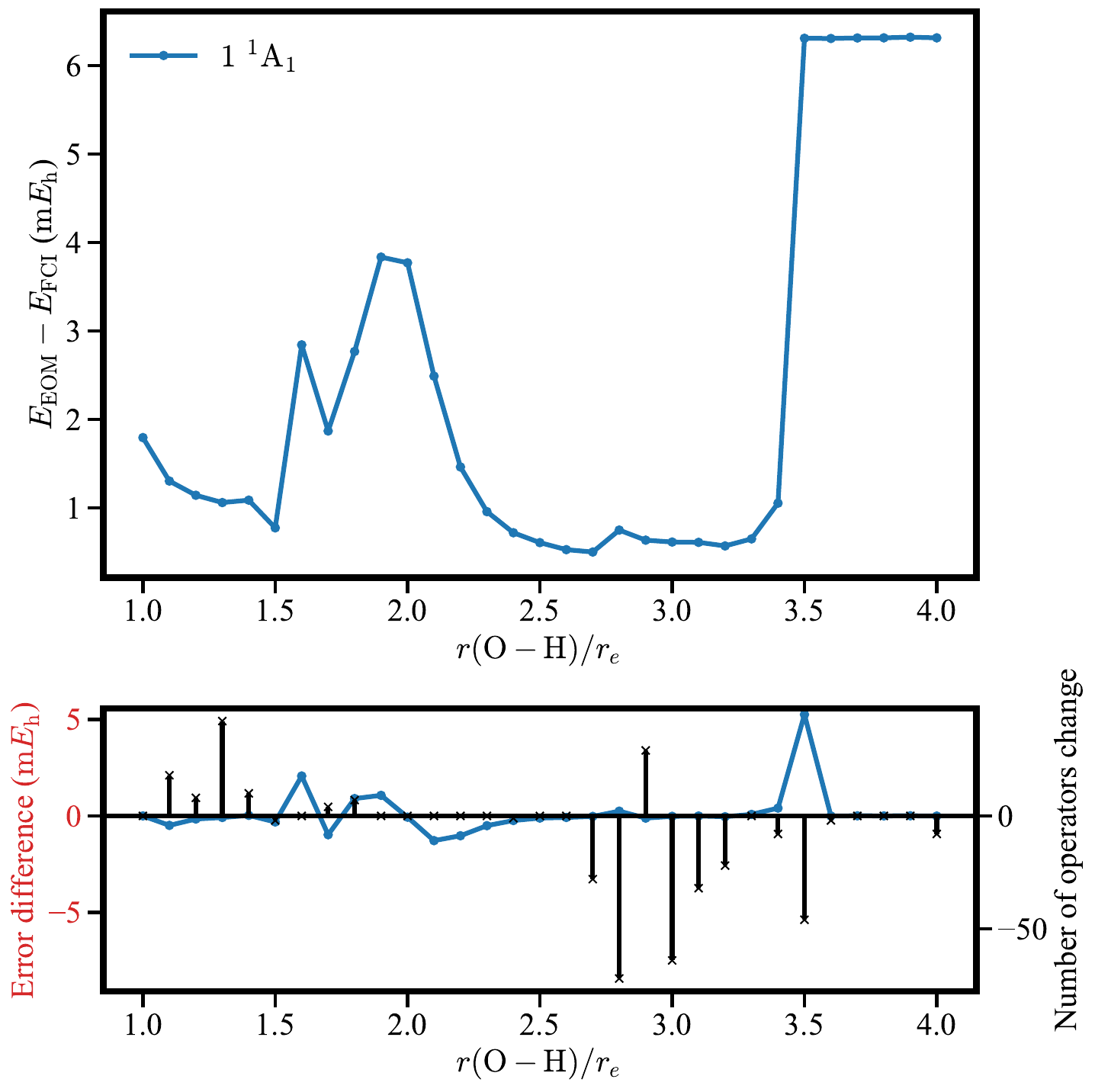} 
\caption{\label{fig:water singlet}
The top panel: potential energy error curves (with respect to FCI) for the $1\ ^1\mathrm{A}_1$ state of the symmetric dissociation of the water computed with the EOM-ic-MRUCCSD method and internal singles and doubles. 
Single threshold $\eta = 10^{-5}$ is used to eliminate linear dependencies.
The lower panel: the change in the number of orthogonal operators and the energy errors with respect to the previous point (to the left).
}
\end{figure}

\begin{figure}[htbp]
\centering
\includegraphics[width = 6.75in]{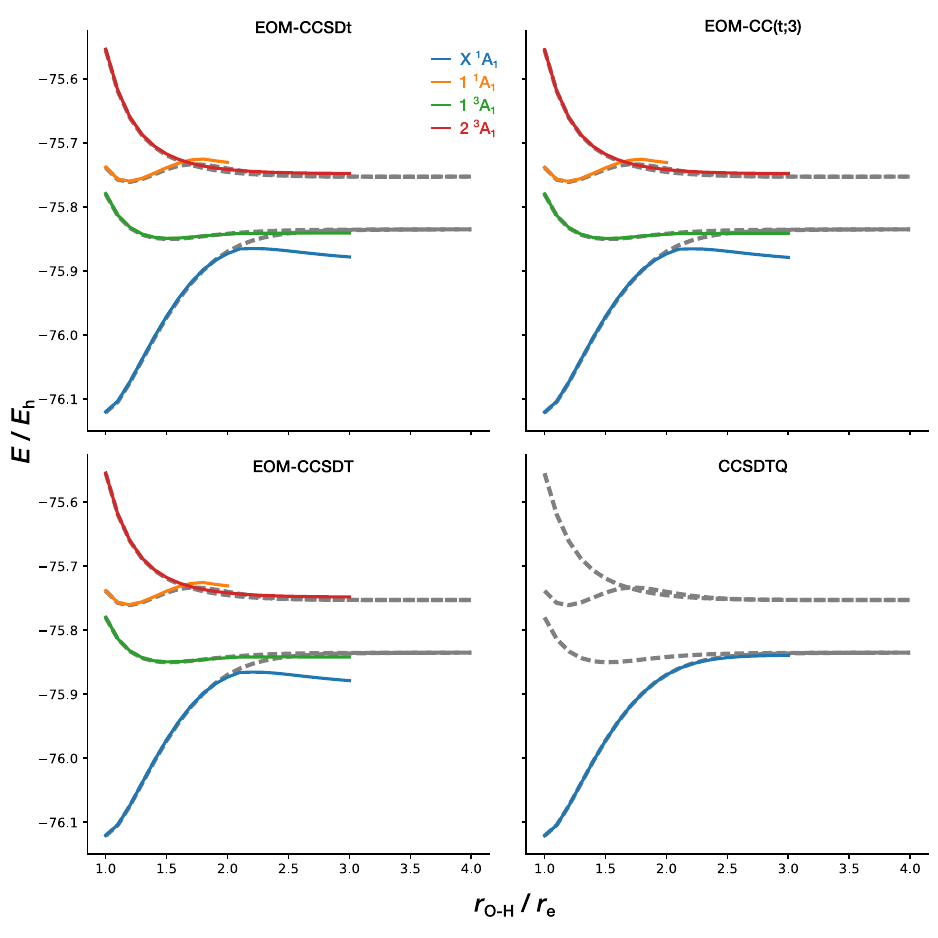} 
\caption{\label{fig:water-compare}
Potential energy curves for the $\mathrm{X}\ ^1\mathrm{A}_1$, $1\ ^1\mathrm{A}_1$, $1\ ^3\mathrm{A}_1$, and $2\ ^3\mathrm{A}_1$ states (colored) of the water symmetric dissociation path computed with different single-reference methods, compared to the FCI results (gray). All methods fail to converge the ground state for $r$(O--H)$>3.0\ r_e$. 
}
\end{figure}

\end{document}